\documentclass[preprint,amsmath,amssymb,aps]{revtex4-2}

\usepackage{graphicx}
\usepackage{dcolumn}
\usepackage{bm}
\usepackage[hidelinks]{hyperref}

\usepackage{subfig}
\usepackage{amsmath,mathtools}
\usepackage{amsthm}
\usepackage{bbold}
\usepackage[shortlabels]{enumitem}
\usepackage{caption}
\usepackage{subcaption}
\usepackage{dcolumn}
\usepackage{bm}
\usepackage{mathrsfs}
\usepackage{xcolor}
\usepackage{float} 

\begin{document}

\title{Model of Simplicial Complexes with dimension wise preferential attachment}

\author{Diego Febbe}
\email{diego.febbe@unifi.it}
\affiliation{Department of Physics and Astronomy, University of Florence \& INFN, Via Sansone
1, 50019 Sesto Fiorentino, Firenze, Italy}
\affiliation{Department of Mathematics \& Namur Institute for Complex Systems - naXys, University of Namur, Rue Joseph Grafé 2, 5000 Namur, Belgium}

\author{Duccio Fanelli}
\affiliation{Department of Physics and Astronomy, University of Florence \& INFN, Via Sansone
1, 50019 Sesto Fiorentino, Firenze, Italy}

\author{Timoteo Carletti}
\affiliation{Department of Mathematics \& Namur Institute for Complex Systems - naXys, University of Namur, Rue Joseph Grafé 2, 5000 Namur, Belgium}


\begin{abstract}
Network science is a powerful framework allowing to model complex systems, capable to describe and take into account the intricate web of connections existing among the constituting basic elements of the system. Recently scholars have brought to the fore the relevance of higher-order networks, namely structures allowing to encode for many-body interactions, differently from the pairwise case handled by networks. This novel research field opens new avenues of research with applications ranging from neuroscience to social sciences and beyond; there is thus a need for generative models of higher-order networks capable to reproduce features present in empirical data. In this work we present a model for growing simplicial complex rooted on a preferential attachment process acting dimension-wise, i.e., returning a power law distribution for the generalized degree of simplexes of different dimension, whose exponent has been analytically determined by studying the growth process and explicitly depends on the probability to add simplices of a given dimension.
\end{abstract}

\maketitle

\section{Introduction}
Over the past several years, complex systems research field has found in network sciences a powerful tool to model interactions among multiple agents \cite{barabasi1999emergence, strogatz2001exploring, albert2002statistical, newman2010networks}. Examples of such versatile descriptions extend across various disciplines, from engineering, biology, and economics to social science \cite{Barabasi2004, nishikawa2015comparative, febbe2024chaos, lucas2023inferring, barabasi2023neuroscience, artime2024robustness, pastor2001epidemic, halaj2024financial, buongiovanni2022will}, to name a few. From a mathematical perspective, algorithms relying on the addition of nodes and links, to eventually build networks encoding complex interactions in large-scale systems have been extensively investigated, leading to the celebrated \textit{preferential attachment} mechanism, introduced and extensively studied in \cite{barabasi1999emergence, dorogovtsev2000evolution, krapivsky2000connectivity, albert2002statistical, bianconi2001competition}. The latter describes the birth and formation of interconnecting structures with a peculiar distribution of node degrees, i.e., power law, that can be found in all the aforementioned real systems.

Supported by new and high-dimensional data, scholars have recently focused on structures that encapsulate higher-order interactions (HOIs), allowing to go beyond the pairwise connections already described by networks. This has led to the development of HOI models, such as simplicial complexes and hypergraphs, which provide a new mathematical framework to study emergent phenomena in complex systems \cite{battiston2020networks, battiston2021physics, bick2023higher, Millan2025}. Applications range from neuroscience, analyzing higher-order organization in neural data \cite{giusti2016two}, to sociology, describing social contagion and activity-driven dynamics \cite{petri2018simplicial, chowdhary2021simplicial, iacopini2019simplicial} to artificial intelligence \cite{morris2019weisfeiler, peri2026spectral, peri2026spectralhigherorderneuralnetworks}. Furthermore, HOI models offer a useful framework for the study of complex dynamics, including the emergence of Turing patterns \cite{muolo2023turing, muolo2024turing}, topological signals \cite{giambagli2022diffusion, calmon2023dirac}, and synchronization phenomena \cite{gambuzza2021stability, gallo2022synchronization, Carletti2023, moriame2025hamiltonian}. Theoretical details and the foundational mathematical tools can be found in comprehensive works covering higher-order networks and algebraic topology related to simplicial complexes \cite{bianconi2021higher, grady2010discrete, lim2020hodge}.

The construction of simplicial complexes with various topologies has been investigated in the literature, with several approaches proposed to model higher-order interactions beyond pairwise networks: from exponential random simplicial complexes \cite{zuev2015exponential}, to the generalization of canonical ensembles \cite{courtney2016generalized}, growing non-equilibrium evolving \textit{flavor} networks, reproducing, among other topologies, preferential attachment \cite{bianconi2016network}, deterministic models \cite{dorogovtsev2025deterministic}, and random simplicial complexes \cite{CostaFarber2016, iacopini2019simplicial}.
 In parallel,  comprehensive collections of real-world datasets are presented in \cite{ahornDataset,hygphx}.

Given the current interest in higher-order topology, and specifically in simplicial complexes (SC), the aim of this work is to propose a model for growing simplicial complexes, where new simplexes are created at each time step. More precisely, with a given probability $p_{d+1}$, a newly added node is used to build a $(d+1)$-simplex, starting from a $d$-simplex randomly chosen among the existing ones, proportionally to its generalized degree. The obtained simplicial complex is thus constructed based on a \textit{preferential attachment} mechanism, depending on the probabilities $p_1,\dots,p_{D}$, where $D$ is the dimension of the simplicial complex. The induced topologies have been studied by analyzing the generalized degree over the growth process, the final degree distribution, and the spectral dimensions, characterizing the convergence rate of dynamical processes defined on them. As we will demonstrate, the resulting topologies are well aligned, under appropriate limits, with established models for the construction of simplicial complexes \cite{wu2015emergent, bianconi2016network, bianconi2017emergent, torres2020simplicial, courtney2016generalized}.

In particular, in \cite{bianconi2016network}, a growing model for the construction of simplicial complexes was introduced, leading to various network topologies ranging from chains and higher-dimensional manifolds to scale-free degree distributions for suitable values of the \textit{flavor} parameter. Here, this paradigm has been expanded by regulating the attachment mechanism through dimension-wise tunable probabilities $p_1, \dots, p_D$, thereby allowing for the introduction of various higher-order structures (links, triangles, tetrahedra, \dots). This allows for the construction of more varied types of simplicial complexes and the study of various topological quantities, such as the exponents governing degree growth or degree distributions, and the spectral dimensions in the mixed regime. See Sec. \ref{sec:topology} for further details and a more concrete comparison with \cite{bianconi2016network}.\\
On the other hand, we would like to highlight that, in \cite{iacopini2019simplicial}, a model regulated by dimension-wise simplex creation was also introduced. However, this model is closer in spirit to Erd\H{o}s--Rényi-inspired random simplicial complexes, keeping the mean node degree fixed in order to enable the study of simplicial contagion.

The work is organized as follows. In Section~\ref{sec:simpcmplx} we present the basic definitions of simplicial complexes needed in the following. The proposed growth algorithm is presented and studied in Section~\ref{sec:topology}. Section~\ref{sec:results} is devoted to the presentation of our results concerning the distribution of the generalized degree and the spectral dimension as a function of the probabilities ${p_{d+1}}$. Finally, in Section~\ref{sec:conclusion}, we draw our conclusions.

We provide a comprehensive Python implementation of the methods and algorithms presented in this work at \cite{diegofebbe_2026_21563618} and it is maintained on \footnote{\url{https://github.com/diegofebbe/Random_walk_on_simplicial_complexes/tree/master}}.


\section{Simplicial Complexes}
\label{sec:simpcmplx}

In this section we present the main properties of simplicial complexes that are useful for the discussion of the following results. A comprehensive theoretical overview of the topic can be found in \cite{bianconi2021higher, grady2010discrete, bick2023higher, lim2020hodge}.

\begin{figure}[!h]
    \centering
    \subfloat[]{\includegraphics[width=0.4\linewidth]{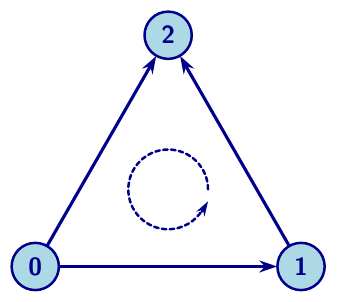}}
    \subfloat[]{\includegraphics[width=0.4\linewidth]{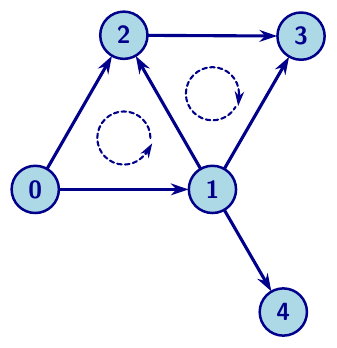}}
    \caption{Representation of a simplex (panel a) and a simplicial complex (panel b) together with their orientations, represented by the arrows on the links and the circular ones inside the triangles. Note that the links orientation is here chosen independently with respect to that induced by the triangles boundary maps. Panel (a): Simplex $[0,1,2]$ (triangle) with the links $[0,1], [1,2]$ coherently oriented with the triangle and $[0,2]$ oppositely oriented. Panel (b): Simplicial complex composed of the simplices $[0,1,2]$, $[1,2,3]$, $[1,4]$ and all their sub-simplices required by the closure property.}
    \label{fig:simplex_theory}
\end{figure}

Let us start by defining a $d$-simplex $\sigma^{(d)} = [i_0, \dots, i_d]$ as the set of $(d+1)$ nodes with the property of containing also all the subsets composed of $\delta\in \{1,\dots,d\}$ unique elements of $\sigma^{(d)}$. Namely the simplex is closed under the inclusion of its faces, i.e., subset of vertices $\{i_0, \dots, i_d\}$. If we consider, for example, the $2$-simplex $[0,1,2]$, i.e., the triangle, shown in Fig.~\ref{fig:simplex_theory}(a), it is composed of the set $\{0,1,2\}$, the sets $\{0,1\}$, $\{0,2\}$, $\{1,2\}$, and the sets $\{0\}$, $\{1\}$, $\{2\}$.
The existence of the triangle $[0,1,2]$ therefore implies the existence of all the sub-simplices obtained by considering subsets of its nodes, namely the three links and the three nodes.

For each $d$-simplex we can define an orientation given by the sign of the permutation $\pi$ of its nodes such that
\begin{equation}
[i_0, \dots, i_d] = (-1)^{p(\pi)}[i_{\pi(0)}, \dots, i_{\pi(d)}],
\end{equation}
where $p(\pi)$ indicates the parity of $\pi$. This allows us to define the boundary map of the $d$-simplex, which returns a linear combination of $(d-1)$-simplices as follows:
\begin{equation}
\label{eq:boundary_map}
\partial_d([i_0, \dots, i_d]) = \sum_{j=0}^d (-1)^j [i_0, \dots, i_{j-1}, i_{j+1}, \dots, i_d]\, .
\end{equation}
By considering the elements of the simplex as basis vectors, we can define the boundary operator $\mathbf{B}_d$ of a simplex $\sigma_j^{(d)}$ as
\begin{equation}
\label{eq:boundary_column_operator}
\mathrm{B}_d(\sigma_i^{(d-1)},\sigma_j^{(d)})=\begin{cases}
    +1 & \text{if $\sigma_i^{(d-1)} \sim \sigma_j^{(d)}$}\\
    -1 & \text{if $\sigma_i^{(d-1)} \not\sim \sigma_j^{(d)}$}\\
    \:\:\:\,0 & \text{otherwise}
\end{cases}\, ,
\end{equation}
where $\sigma_i^{(d-1)} \sim \sigma_j^{(d)}$ means that $\sigma_i^{(d-1)}$ is a face of $\sigma_j^{(d)}$ and they are coherently oriented, meaning that the independent lexicographic orientation of $\sigma_i^{(d-1)}$ coincides with that induced by the boundary map applied to $\sigma_j^{(d)}$, while $\sigma_i^{(d-1)} \not\sim \sigma_j^{(d)}$ denotes the fact that $\sigma_i^{(d-1)}$ is a face of $\sigma_j^{(d)}$ and they are not coherently oriented.
Let us observe that in the following we will also use the lightened notation $B_d(i,j)\equiv \mathrm{B}_d(\sigma_i^{(d-1)},\sigma_j^{(d)})$.
As an example, we can compute the operators $\mathbf{B}_1$ an $\mathbf{B}_2$ of the simplex $\sigma^{(2)}=[0,1,2]$ shown in Fig.~\ref{fig:simplex_theory}(a):
\begin{equation}
\mathbf{B}_1 = \bordermatrix{
& \scriptstyle[0,1] & \scriptstyle[0,2]& \scriptstyle[1,2]\cr
\scriptstyle[0] & -1 & -1 & 0\cr
\scriptstyle[1] & 1 & 0 & -1\cr
\scriptstyle[2] & 0 & 1 & 1\cr
}\quad\text{ and }\quad
\mathbf{B}_2 = \bordermatrix{
& \scriptstyle[0,1,2] \cr
\scriptstyle[0,1] & 1 \cr
\scriptstyle[0,2] & -1 \cr
\scriptstyle[1,2] & 1 \cr
}\, .
\end{equation}
For instance, for the oriented simplex $\sigma^{(2)}=[0,1,2]$ we have
$\partial_2[0,1,2]=[1,2]-[0,2]+[0,1]$. Therefore, if the edge basis is chosen by lexicographical order, the faces $[0,1]$ and $[1,2]$ are coherently oriented with $[0,1,2]$, whereas $[0,2]$ is not coherently oriented due to the minus sign. This is highlighted in Fig. \ref{fig:simplex_theory}(a) where the arrow $[0,2]$ opposes the triangle orientation (internal, circular arrow) giving $(1,-1,1)^\top$ for $\mathbf{B}_2$.
This definition can be generalized to a $D$-simplicial complex, $\mathcal{X}$, obtained by gluing together $N$ simplexes (see Fig.~\ref{fig:simplex_theory}(b) for a simple example, whose boundary operators are shown in Eq.~\eqref{eq:B1B2simplexFig1b}).
\begin{equation}
\label{eq:B1B2simplexFig1b}
\begin{aligned}
\mathbf{B}_1 = \bordermatrix{
& \scriptstyle[0,1] & \scriptstyle[0,2] & \scriptstyle[1,2] & \scriptstyle[1,3] & \scriptstyle[1,4] & \scriptstyle[2,3] \cr
\scriptstyle[0] & -1 & -1 &  0 &  0 &  0 &  0 \cr
\scriptstyle[1] &  1 &  0 & -1 & -1 & -1 &  0 \cr
\scriptstyle[2] &  0 &  1 &  1 &  0 &  0 & -1 \cr
\scriptstyle[3] &  0 &  0 &  0 &  1 &  0 &  1 \cr
\scriptstyle[4] &  0 &  0 &  0 &  0 &  1 &  0 \cr
}
\quad\text{ and }\quad
\mathbf{B}_2 = \bordermatrix{
& \scriptstyle[0,1,2] & \scriptstyle[1,2,3] \cr
\scriptstyle[0,1] & 1 & 0 \cr
\scriptstyle[0,2] & -1 & 0 \cr
\scriptstyle[1,2] & 1 & 1 \cr
\scriptstyle[1,3] & 0 & -1 \cr
\scriptstyle[1,4] & 0 & 0 \cr
\scriptstyle[2,3] & 0 & 1 \cr
}\, .
\end{aligned}
\end{equation}

By using the boundary operator, we can generalize the concept of the Laplace matrix to higher-order structures, and thus define the $n$-dimensional Hodge-Laplace operator~\cite{torres2020simplicial, giambagli2022diffusion}
\begin{equation}
\label{eq:Ln}
\mathbf{L}_d = \mathbf{L}_d^{\mathrm{down}} + \mathbf{L}_d^{\mathrm{up}}\, ,
\end{equation}
with
\begin{equation}
\mathbf{L}_d^{\mathrm{down}}  = \mathbf{B}_d^\top \mathbf{B}_d \text{ and }
\mathbf{L}_d^{\mathrm{up}} = \mathbf{B}_{d+1} \mathbf{B}_{d+1}^\top\, .
\end{equation}
Let us observe that $\mathrm{L}_d^{\mathrm{down}}(i,j)\neq 0$ denotes the fact that the $d$-simplexes, $\sigma^{(d)}_i$ and $\sigma^{(d)}_j$, share a common face, i.e., a $(d-1)$-simplex. On the other hand  $\mathrm{L}_d^{\mathrm{up}}(i,j)\neq 0$ implies that $\sigma^{(d)}_i$ and $\sigma^{(d)}_j$ are included into the same $(d+1)$-simplex.

Let us observe that the usual (combinatorial) graph Laplacian
\begin{equation}
\mathbf{L}_0 = \mathbf{D}_{N_0} - \mathbf{A}\, ,
\end{equation}
where $\mathbf{D}_{N_0}$ denotes the diagonal matrix with the node degrees on the diagonal and $\mathbf{A}$ is the graph adjacency matrix, can be obtained as a special case of~\eqref{eq:Ln} once we set $\mathbf{B}_0=0$, because the simplex does not contain any object smaller than nodes.

We now present some properties of the spectrum of the $d$-Laplace matrix. The nonzero spectrum of $\mathbf L_d$ is given by the union of the nonzero
spectra of $\mathbf L_d^{\mathrm{up}}$ and $\mathbf L_d^{\mathrm{down}}$.
Let us consider an eigenvector $\mathbf v$ of $\mathbf L_d^{\mathrm{up}}$
associated with a nonzero eigenvalue $\lambda$, namely
\begin{equation}
\lambda \mathbf{v} = \mathbf{L}^\mathrm{up}_d \mathbf{v} = \mathbf{B}_{d+1} \mathbf{B}_{d+1}^\top\mathbf{v}\, ,
\end{equation}
by applying the down-Laplacian we obtain
\begin{equation}
\lambda \mathbf{L}^\mathrm{down}_d\mathbf{v} = \lambda \mathbf{B}_{d}^\top \mathbf{B}_{d}  \mathbf{v} =  \mathbf{B}_{d}^\top \mathbf{B}_{d} \mathbf{B}_{d+1} \mathbf{B}_{d+1}^\top \mathbf{v}= 0\, ,
\end{equation}
since $\mathbf{B}_d \mathbf{B}_{d+1} = 0$; being $\lambda>0$ we can conclude that $\mathbf{v}\in \ker \left(\mathbf{L}^\mathrm{down}_d\right)$. The remaining case can be handled in a very similar way.

Furthermore, $\mathbf{L}^\mathrm{down}_{d+1}$ and $\mathbf{L}^\mathrm{up}_d$ share the same spectrum, except for the multiplicity of the zero eigenvalue. Indeed, by considering the singular value decomposition 
\begin{equation}
\mathbf{B}_{d+1} = \mathbf{U} \Sigma \mathbf{V}^\top\, ,
\end{equation}
where $\mathbf{U} \in \mathbb{R}^{N_{d} \times N_{d}}$ and $\mathbf{V} \in \mathbb{R}^{N_{d+1} \times N_{d+1}}$ are orthogonal matrices, and $\Sigma \in \mathbb{R}^{N_{d} \times N_{d+1}}$ is a diagonal rectangular matrix containing the singular values of $\mathbf{B}_{d+1}$. Then by direct computation we obtain
\begin{equation}
\begin{aligned}
\mathbf{B}_{d+1} \mathbf{B}_{d+1}^\top
&= \mathbf{U} \Sigma \mathbf{V}^\top \mathbf{V} \Sigma^\top \mathbf{U}^\top
= \mathbf{U} \Sigma \Sigma^\top \mathbf{U}^\top \text{ and }\\
\mathbf{B}_{d+1}^\top \mathbf{B}_{d+1}
&= \mathbf{V} \Sigma^\top \mathbf{U}^\top \mathbf{U} \Sigma \mathbf{V}^\top
= \mathbf{V} \Sigma^\top \Sigma \mathbf{V}^\top\, .
\end{aligned}
\end{equation}
It is straightforward to note that $\Sigma \Sigma^\top$ and $\Sigma^\top \Sigma$ share the same nonzero spectrum, and this property is preserved under the orthogonal transformations $\mathbf{U}$ and $\mathbf{V}$.

Moreover, the $d$-Laplacian is positive semi-definite, implying that its spectrum ${\lambda_\ell}$ satisfies
\begin{equation}
0 \leq \lambda_1 \leq \lambda_2 \leq \dots \leq \lambda_{N_d}\, ,
\end{equation}
where the multiplicity of the zero eigenvalue is equal to the corresponding Betti number $\mathfrak{b}_d$~\cite{torres2020simplicial, bianconi2017emergent}.

By considering the density of eigenvalues $\rho(\lambda)$, for $\lambda \ll 1$, we can define the \textit{spectral dimension} $d_s^{(0)}$ as
\begin{equation}
\rho(\lambda) \simeq \tilde{C}_{(0)} \lambda^{\frac{d_s^{(0)}}{2} - 1},
\end{equation}
or, equivalently, through its cumulative distribution $\rho_c(\lambda)$,
\begin{equation}
\rho_c(\lambda) \simeq C_{(0)} \lambda^{d_s^{(0)}/2}.
\end{equation}
We can generalize the spectral dimension to a generic (topological) dimension $d>0$ by considering the spectrum of the up-Laplacian $\mathbf{L}_d^{\mathrm{up}}$ \cite{torres2020simplicial}, and obtaining thus 
\begin{equation}
\rho_c^{\mathrm{up}}(\lambda)_d \simeq C_{(d)} \lambda^{d_s^{(d)}/2}\, .
\end{equation}

The spectral dimension, defined through the distribution of the Laplacian eigenvalues, provides a measure of the structural dimension as experienced by a diffusion process \cite{millan2018complex, millan2019synchronization, millan2020explosive, torres2020simplicial, burioni2004topological}.
Furthermore, for $\lambda \ll 1$ \cite{torres2020simplicial} we can write 
$$
\rho_c = \frac{\#\{\lambda'<\lambda\}}{N_0},
$$
and by posing $\lambda = \lambda_2$,
$$\#\{\lambda'<\lambda_2\} \simeq  N_0C_0\lambda_2^{d_s(0)/2},
$$
leading to
\begin{equation}
\label{eq:lambda_2_scaling}
\lambda_2 \propto N_0^{-2/d_s^{(0)}},
\end{equation}
which provides a reference for the convergence time of the diffusion process regulated by the Fiedler eigenvalue $\lambda_2$. If $\beta_d=1$, Eq. (17) can be generalized as
\[
\lambda_2 \propto N_d^{-2/d_s^{(d)}}.
\]
In addition, in some cases the spectral dimension is also related to the stability properties of synchronized oscillators, as in the Kuramoto model \cite{torres2020simplicial}.

\section{The algorithm for the growing simplicial complex} 
\label{sec:topology}

Our work is loosely inspired by the framework introduced in \cite{bianconi2016network}, where different topological constructions of simplicial complexes are systematically explored. Among the presented methods, with an appropriate \textit{flavor} value $s=1$, a parameter that modulates the network construction, a \textit{preferential attachment} mechanism for building simplicial complexes of dimension $D$ is introduced. The proposed growing process assumes that at each step a new node is added to the network, by forming a new $D$-dimensional simplex  together with all the lower-dimensional simplices required by the closure property. The $D$-simplex is formed by the newly added node and a $(D-1)$-face of an existing simplex, chosen proportionally to its generalized degree.

Leveraging this framework, we hereby present a method for simplicial complex generation rooted in a dimension-wise \textit{preferential attachment} mechanism. To compute the generalized degree and its evolution as the process evolves, we introduce the ``unsigned'' incidence matrix $\mathbf{M}_d=|\mathbf{B}_d|$, where the absolute value function applies entry-wise, namely
\begin{equation}
    \label{eq:not_normalized_M}
    \mathrm{M}_d(i,j)=|\mathrm{B}_d(i,j)|\, .
\end{equation}
Let us observe that the latter matrix has been recently used to define a random walk process on simplicial complex, where the walker can hop between simplexes whose dimension differs by one unity~\cite{febbe2026random}.

Given a $(d-1)$-simplex, $\sigma_i^{(d-1)}$, we can define its ``upper'' degree
\begin{equation}
\label{eq:dsigmaup}
k_{{up}_{i}}^{(d-1)} = \sum_{j=1}^{N_d} \mathrm{M}_d(\sigma_i^{(d-1)},\sigma_j^{(d)})\, ,
\end{equation}
namely the number of $d$-simplices to which it belongs. We can also introduce the ``lower'' degree, $k_{{down}_{j}}^{(d)}$, of the $d$-simplex, $\sigma_j^{(d)}$, namely the number of $(d-1)$-simplices contained into it; however because of the inclusion closure property of the simplicial complex, we trivially have
\begin{equation}
\label{eq:dsigmadwn}
k_{{down}_{j}}^{(d)} = \sum_{i=1}^{N_{d-1}} \mathrm{M}_d(\sigma_i^{(d-1)},\sigma_j^{(d)})\equiv d+1\, .
\end{equation}
Let us observe that $k_{{down}_{j}}^{(d)}$ is defined for $d\geq 1$, while $k_{i}^{(d-1)}$ for $d\leq D$ since 0 and $D$ are, respectively, the minimum and maximum allowed dimensions.

Let us now present in detail the algorithm for the growing simplicial complex. At each time step a new node enters and with probability $p_{d+1}$ it creates a new simplex of dimension $(d+1)$, clearly $\sum_{d=0}^{D-1} p_{d+1} = 1$. The latter structure is formed by merging the entering nodes with an existing $d$-simplex, the latter will be drawn with probability proportional to its degree.

For the sake of simplicity, the analysis presented throughout this work (with exception of Appendix~ \ref{sec:Lower degree included}) has been conducted by considering only the upper degree $k_{{up}_i}^{(d)}$ given by Eq.~\eqref{eq:dsigmaup}, instead of the total degree $K_i^{(d)} = k_{{up}_i}^{(d)} + k_{{down}_{i}}^{(d)}$. For this reason, for simplicity of exposition, from now on we will denote the upper degree of a simplex $\sigma_i^{(d)}$ just as $k_i^{(d)}$ (with the exception of Appendix~ \ref{sec:Lower degree included}). 
Accordingly, the probability of selecting a given existing simplex $\sigma_i^{(d)}$ is given by
\begin{equation}
\Pi_i^{(d)} = \frac{k_{i}^{(d)}}{\sum_j k_{j}^{(d)}}\, .
\label{eq:probability_with_uppdegree}
\end{equation}
Let us recall that the lower degree $k_{{down}_{i}}^{(d)}$, given by Eq.~\eqref{eq:dsigmadwn}, is constant once the dimension is fixed, therefore, including it in the selection probability would introduce a bias toward higher dimensions. This choice is, however, by no means mandatory and similar analyses can be carried out equally by also considering the lower degree (see Appendix~ \ref{sec:Lower degree included}).
Moreover, let us notice, that with the parameter definition given by Eq. \eqref{eq:probability_with_uppdegree}, in cases with $p_D=1$ (and thus $p_d=0$ for $d=1,\dots,D-1$), the proposed algorithm returns the method proposed in~\cite{bianconi2016network} with \textit{flavor} parameter set to $s=1$. It is also important to note that, to ensure the proposed process is meaningful, an initial seed with some simplices having $k^{(d)}>0$ is required to attach higher-order structures.

\begin{figure}[!h]
    \centering
    \includegraphics[width=0.9\linewidth]{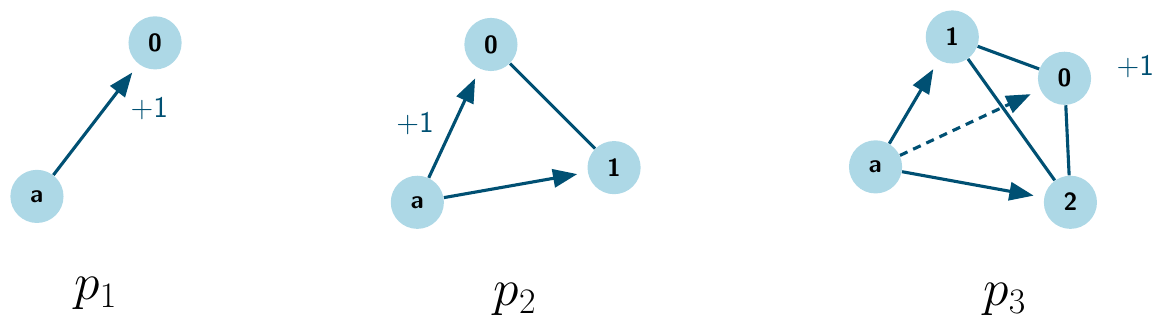}
    \caption{Growing process for the simplicial complex. With probability $p_1$, the incoming node forms a link by connecting to an already present node. With probability $p_2$, a (filled) triangle is formed by connecting the incoming node to a link, while with probability $p_3$ a new (filled) tetrahedron is formed, and so on. Note that the degree of the already present nodes always increases by 1 in all cases, by connecting, respectively, to a newly formed link.}
    \label{fig:simplex_growth}
\end{figure}

Before presenting the algorithm in its full generality, let us consider some simple examples shown in Fig.~\ref{fig:simplex_growth}. Once the new node $a$ is added to the simplex, with probability $p_1$ it will form a $1$-simplex, namely a link; to do this we select among the existing $0$-simplices, i.e., the nodes, the one with whom $a$ will form the new link, this node will thus be selected according to $k_i^{(0)}/\sum_j k^{(0)}_j$. Namely, the classical preferential attachment proposed by Barab\'asi and Albert~\cite{barabasi2016network}. To go beyond this case, we assume the entering node to form a $2$-simplex, i.e., a triangle, with probability $p_2$. The former must thus attach to any existing $1$-simplex, i.e., a link, the latter being selected according to $k_i^{(1)}/\sum_j k^{(1)}_j$. This idea can be generalized to higher-dimension simplexes, for instance with probability $p_3$, node $a$ will merge with an existing $2$-simplex, a triangle, selected with probability $k_i^{(2)}/\sum_j k^{(2)}_j$, to form a $3$-simplex, a tetrahedron.

We can now present the proposed algorithm in its full generality. The process consists of three consecutive steps that will be iterated as long as required to reach the sought size of the simplicial complex (normally given in terms of number of nodes $N_0$):
    \begin{itemize}
    \item A new node, say $a$, is added to the simplicial complex;
    \item With probability $p_{d+1}$, $d\in\{0,\dots,D-1\}$, we select the dimension of the simplex to be formed;
    \item A simplex $\sigma_j^{(d)}$ of dimension $d$ is selected among the existing ones, with probability $\Pi_j^{(d)} = k_j^{(d)}/\sum_h k_h^{(d)}$. The latter will serve as building block to form a $(d+1)$-simplex by adding the node $a$. Of course, all the sub-simplices will also be added, as required by the closure property.
\end{itemize}
This algorithm provides a general method for generation of simplicial complexes with tunable parameters $p_{d+1}$ defining different \textit{dimension-wise preferential attachment} topologies.

\section{Results}
\label{sec:results}

The aim of this section is to present our results to describe some topological features of the simplicial complexes built by using the algorithm presented in the previous section. In particular we will analytically determine the ``time'' evolution of the generalized degree and thus its asymptotic distribution, and we will then numerically study the spectral dimension. In both cases, the main parameters under control will the probabilities $p_{d+1}$.

\subsection{On the evolution of the degree}
\label{ssec:degevol}

By following~\cite{barabasi2016network} we can determine an equation to describe the generalized degree evolution by approximating the discrete construction as a continuous process. Let us observe that a similar result can be obtained by analyzing the asymptotic behavior of the discrete growing process (see Appendix~\ref{sec:appendix_discrete_version}). 

Given a simplicial complex of dimension $D$, the evolution of the (upper) degree of a simplex $\sigma_i^{(d)}$ is given by:
\begin{align}
\frac{d k_i^{(d)}}{dt} &= p_{d+1} \Pi_i^{(d)} + p_{d+2} \sum_{j} \mathrm{M}_{d+1}(i,j) \Pi_j^{(d+1)} + p_{d+3} \sum_{j_1,j_2} \frac{\mathrm{M}_{d+1}(i,j_1) \mathrm{M}_{d+2}(j_1,j_2)}{2!}\Pi_{j_2}^{(d+2)} \nonumber \\
& \quad + \dots + p_D\sum_{j_1, \dots,j_{D-d-1}} \frac{\mathrm{M}_{d+1}(i,j_{1}) \dots \mathrm{M}_{D-1}(j_{D-d-2},j_{D-d-1})}{(D-d-1)!}\Pi_{j_{D-d-1}}^{(D-1)}\, ,
\label{eq:degree_general_equation}
\end{align}
the leftmost term on the right hand side, denotes the process of the creation of a $(d+1)$-simplex by selecting the $i$-th $d$-simplex with probability $\Pi_i^{(d)}$ and by adding the new entering node, hence $k_i^{(d)}$ will increase. The next term represents the creation of a $(d+2)$-simplex by selecting a $(d+1)$-simplex, say $j$, and by taking into account that $i$ should be incident with $j$, namely $\mathrm{M}_{d+1}(i,j)=1$. The remaining terms encode similar processes where each time we have to check that the inclusion property is satisfied.

The factorial at the denominators are normalizing terms, formally given by
\begin{eqnarray}
    \label{eq:normfact}
    2! &=& \sum_{j_1}\mathrm{M}_{d+1}(i,j_1) \mathrm{M}_{d+2}(j_1,j_2)\\
    3!&=&\sum_{j_1j_2}\mathrm{M}_{d+1}(i,j_1) \mathrm{M}_{d+2}(j_1,j_2)\mathrm{M}_{d+3}(j_2,j_3)\notag \\
    &\vdots& \notag \\
    (D-d-1)!&=&\sum_{j_1,\dots,j_{D-d-2}}\mathrm{M}_{d+1}(i,j_1) \dots \mathrm{M}_{D-1}(j_{D-d-2},j_{D-d-1})\notag\, ,
\end{eqnarray}
if simplex $\sigma_i^{(d)}$ is a face contained in the highest-order simplex of the chain.
We hereby provide the intuition behind those relations, while the interested reader can find a formal proof in Appendix~\ref{sec:number_of_paths}. Let us consider the first of equations~\eqref{eq:normfact}; let $\sigma_i^{(d)}=[i_0,\dots,i_d]$ be an oriented $d$-simplex and let $\sigma_{j_2}^{(d+2)}$ be an oriented $(d+2)$-simplex containing $\sigma_i^{(d)}$, hence it should be of the form $\sigma_{j_2}^{(d+2)}=[i_0,\dots,i_d,\ell_1,\ell_2]$ for some nodes $\ell_2>\ell_1>i_d$. This leaves only two possibilities for a $(d+1)$-simplex $\sigma_{j_1}^{(d+1)}$: either $[i_0,\dots,i_d,\ell_1]$ or $[i_0,\dots,i_d,\ell_2]$.

As an example, we note in Fig.~\ref{fig:simplex_growth} that the degree of node $0$ always increases by $1$ when connecting to a new link, even in the rightmost scenario, where a new tetrahedron is formed by using the existing triangle $[0,1,2]$, to which node $0$ belongs, as the basis for the connection.

\begin{figure}[h!!]
    \centering
    \subfloat[]{\includegraphics[width=0.45\linewidth]{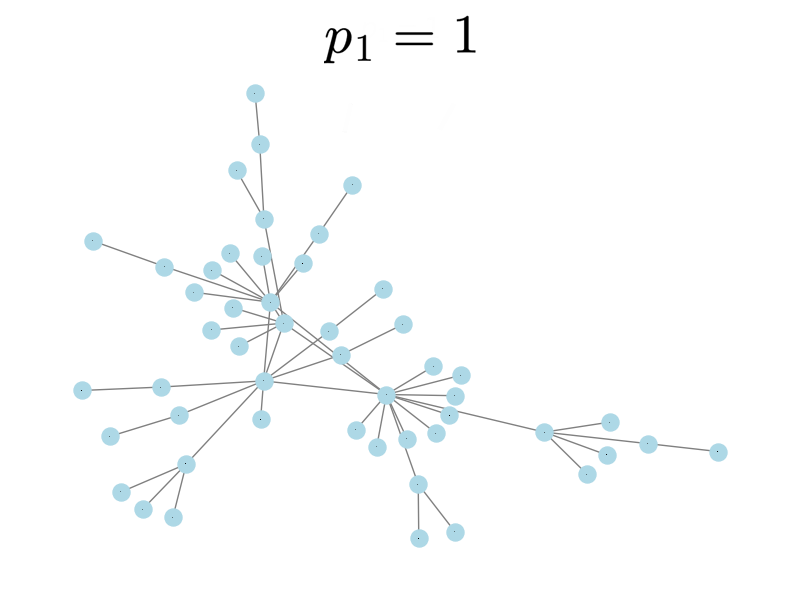}}
    \subfloat[]{\includegraphics[width=0.45\linewidth]{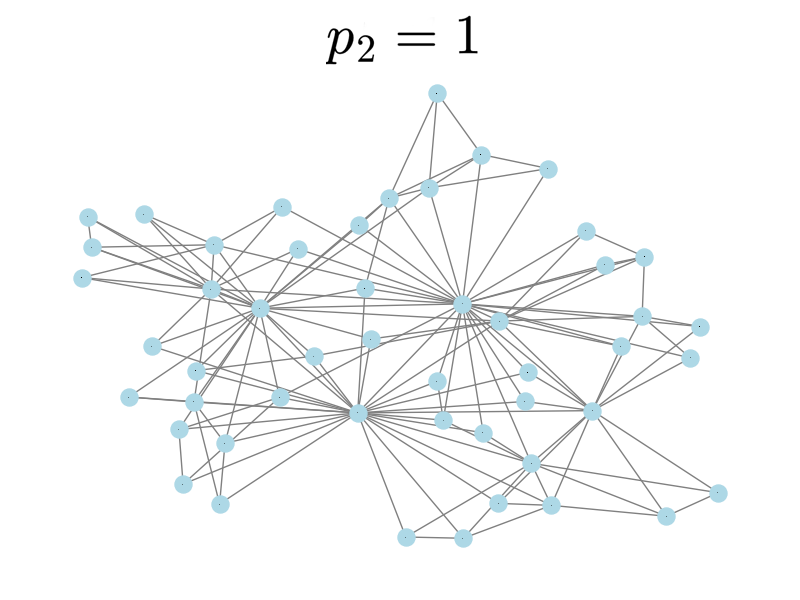}}\\
    \subfloat[]
    {\includegraphics[width=0.45\linewidth]{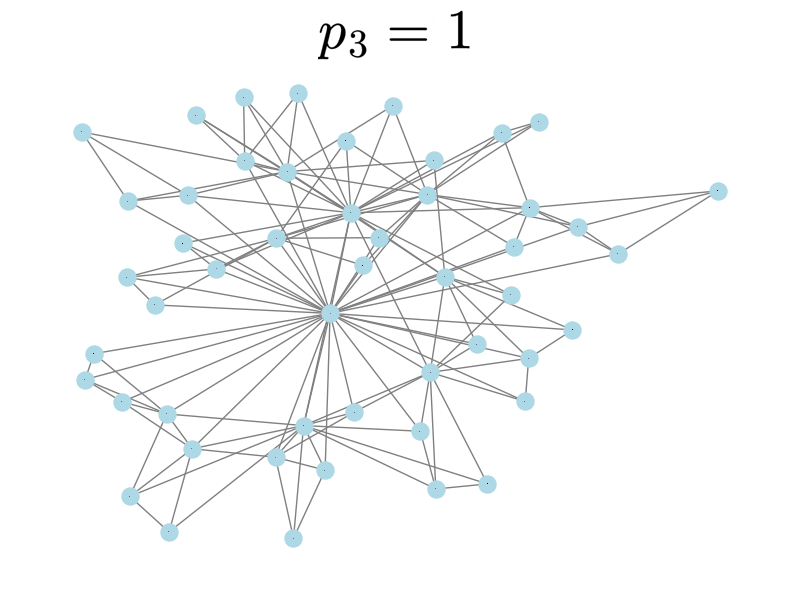}}
    \subfloat[]
    {\includegraphics[width=0.45\linewidth]{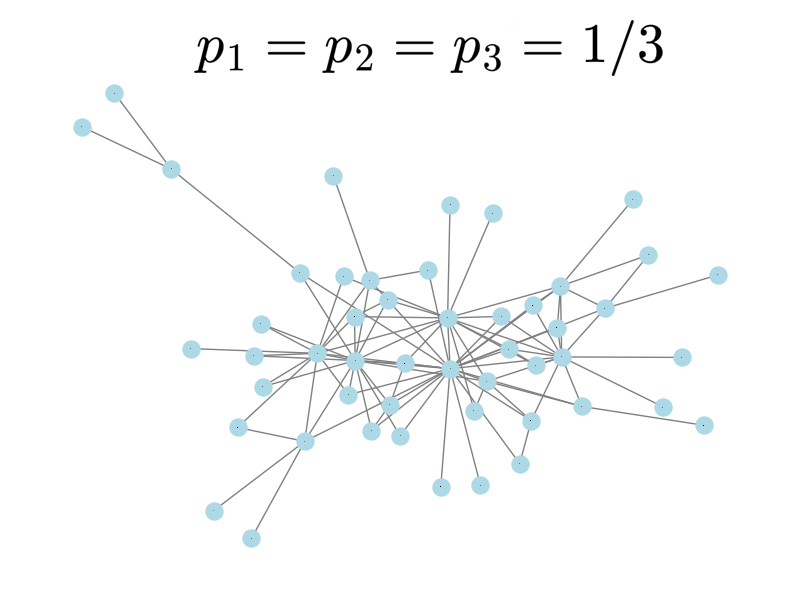}}
    \caption{Examples of various simplex generations, by varying the parameters $p_1$, $p_2$ and $p_3$. In all these cases the number of nodes is fixed to $N_0=50$ and the seed simplicial complex is a tetrahedron, together with the 4 triangles, 6 links, and 4 nodes required by the closure property. The structure created in the subsequent $N_{\mathrm{steps}}=46$ steps of the algorithm can be computed analytically for the first three cases: $N_{\mathrm{steps}}$ nodes and links in panel (a); $N_{\mathrm{steps}}$ nodes, $2N_{\mathrm{steps}}$ links, and $N_{\mathrm{steps}}$ triangles in panel (b); and $N_{\mathrm{steps}}$ nodes, $3N_{\mathrm{steps}}$ links, $3N_{\mathrm{steps}}$ triangles, and $N_{\mathrm{steps}}$ tetrahedra in panel (c). For the mixed case, shown in panel (d), we compute the average added structure as $N_{\mathrm{steps}}$ nodes, $(p_1+2p_2+3p_3)N_{\mathrm{steps}}$ links, $(p_2+3p_3)N_{\mathrm{steps}}$ triangles, and $p_3N_{\mathrm{steps}}$ tetrahedra. Here, higher-order interactions are not represented in order to keep the image uncluttered and visually clear.
    } 
    \label{fig:various_simplex_plots}
\end{figure}

For a generic choice of the building parameters ${p_{d+1}}$, at each (building) time step, simplices of different dimensions can be formed, creating simplicial complexes composed of heterogeneous structures. Examples of this are shown in Fig. \ref{fig:various_simplex_plots}, where different simplicial complexes have been created and shown for some choice of the building parameters ${p_{d+1}}$, while keeping the seed simplex, to which new simplices attach, fixed. By varying the probabilities ${p_{d+1}}$, different types of higher-order structures, such as links, triangles, tetrahedra, and so on, can be included in the simplicial complex, together with all the sub-simplices required by the downward closure. These structures, although not explicitly represented for clarity, are assumed to be complete in these simulations, according to \cite{bianconi2016network}, not allowing the construction of "empty" structures (such as the three links without the corresponding filled triangle). Different variants of this generation algorithm, are discussed in Appendix \ref{sec:mixed_probabilities_computation}.

\begin{figure}
    \centering
    \includegraphics[width=0.8\linewidth]{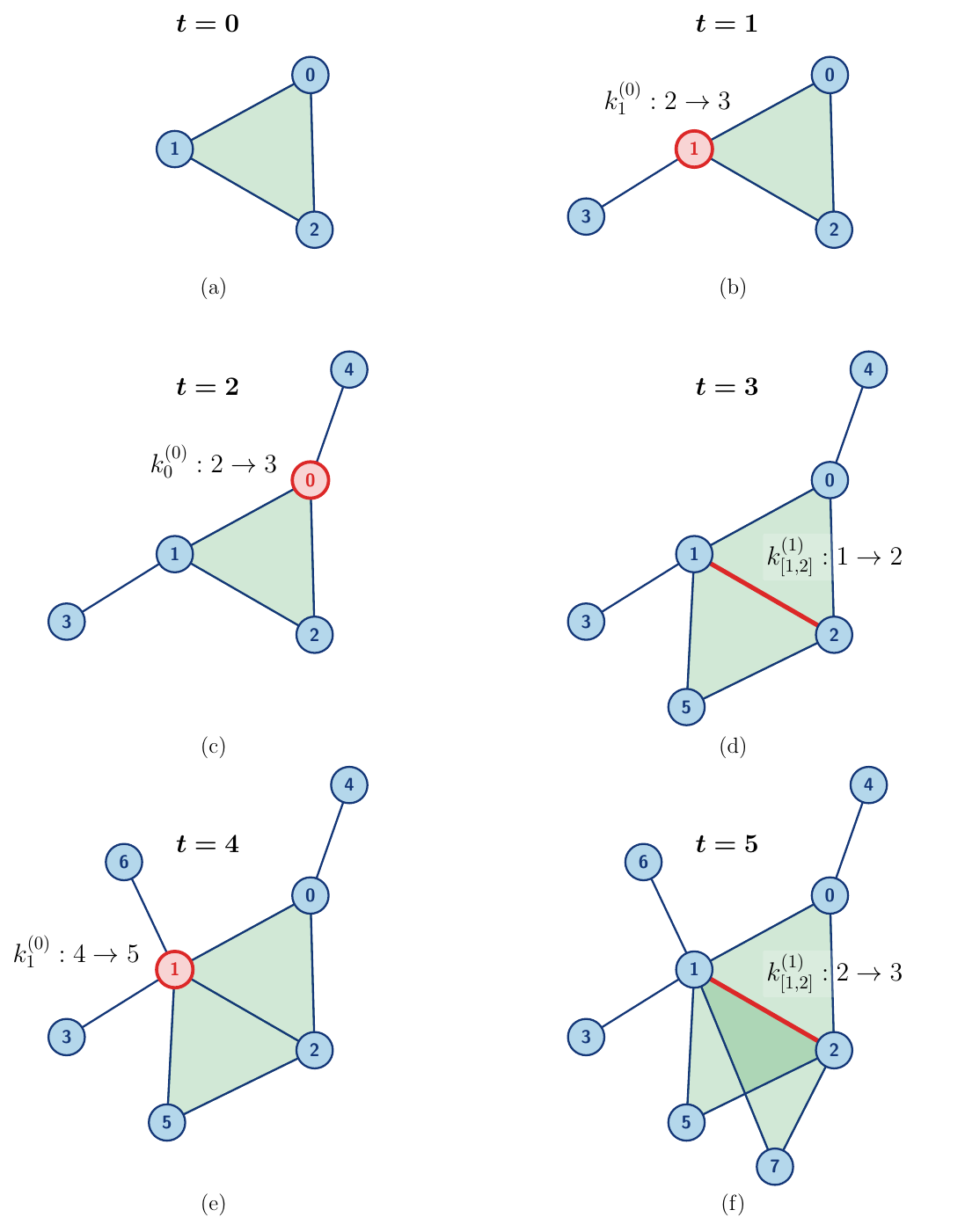}
    \caption{Example of the growth process of a simplicial complex with $p_1=p_2=\frac{1}{2}$, starting from the triangle seed $[0,1,2]$ (panel (a)). At the first algorithmic time step (panel (b)), a link-creation event, modulated by the parameter $p_1$, is selected, and a node is chosen with probability proportional to its degree; here, node $1$ whose (up-)degree increases $(2 \to 3)$. The same type of event occurs in panels (c) and (e). In panels (d) and (f), instead, a triangle-creation event, modulated by $p_2$, is selected; in panel (d), the new triangle is attached to link $[0,1]$, whose (up-)degree increases $(1 \to 2)$. Due to the closure property, the degrees $k_1^{(0)}$ and $k_1^{(2)}$ also increase, since the nodes [1] and [2] attach to the newly created links.}
    \label{fig:simplicial_complex_simple_evolution}
\end{figure}
In Fig. \ref{fig:simplicial_complex_simple_evolution}, we instead report an example of the growth process over a few time steps, illustrating the basic generation mechanism. Here, we set $p_1 = p_2 = \frac{1}{2}$ for visual clarity, also highlighting the presence of higher-order interactions (represented by the filled triangles). In this example, simplices can enter through the two channels regulated by $p_1$ and $p_2$ and attach to already existing structures with probability proportional to their base degree.

By following \cite{barabasi2016network}, in the case $p_D = 1$, and thus $p_d=0$ for all $d=0,\dots,D-1$, Eq.~\eqref{eq:degree_general_equation} can be explicitly solved (see Appendix~\ref{sec:power_law_exponent_derivation}), by yielding the following expression for the degree evolution of a simplex $i$, which joins the network at time $t_i$, as a function of the building time $t$:
\begin{equation}
k^{(d)}_i(t) = \left(\frac{t}{t_i}\right)^{\beta_d} \, ,
\label{eq:degree_evolution_beta}
\end{equation}
where $\beta_d = \frac{D-d}{D+1}$. Hence, the distribution of the ``upper'' degree of simplices of dimension $d$ follows a power law with exponent $\gamma_d = 1 + \frac{1}{\beta_d}$
\begin{equation}
\label{eq:gamma_power_law}
\gamma_d = 1 + \frac{1}{\beta_d} = 1 + \frac{D+1}{D-d}\, .
\end{equation}

\begin{figure}[h!!]
\centering
\subfloat[]{\includegraphics[width=0.45\linewidth]{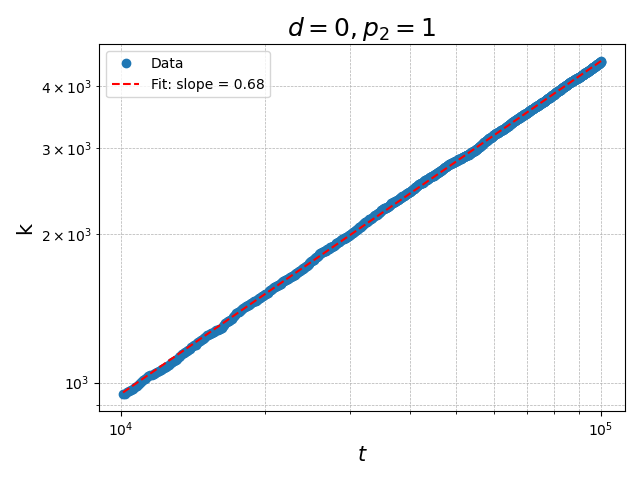}}
\subfloat[]{\includegraphics[width=0.45\linewidth]{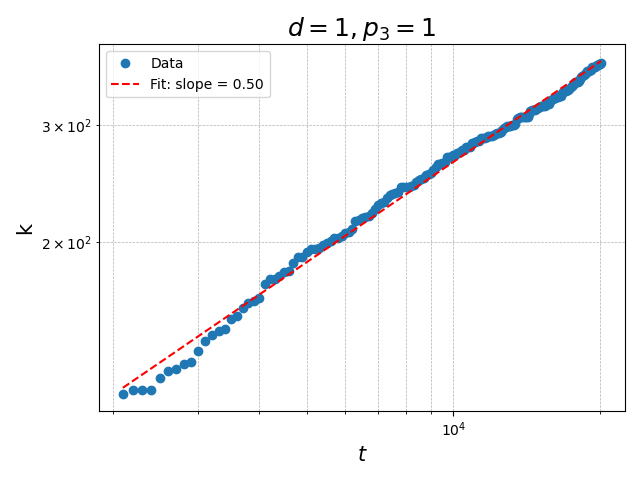}}\
\subfloat[]
{\includegraphics[width=0.45\linewidth]{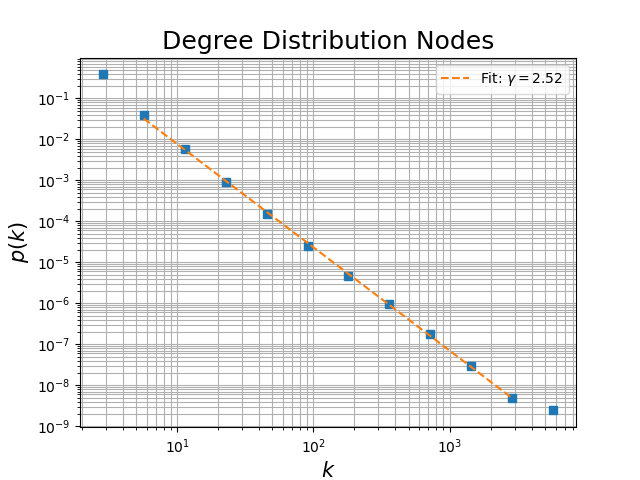}}
\subfloat[]
{\includegraphics[width=0.45\linewidth]{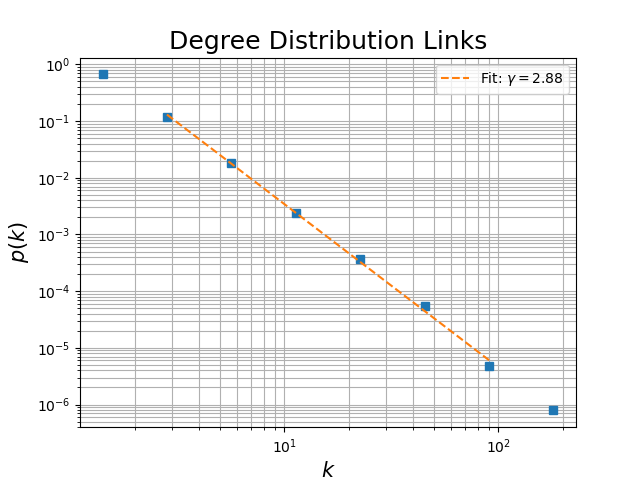}}
\caption{Numerical simulation results for the values of $\beta_d$ and $\gamma_d$ are shown for $d=0$, $p_2=1$ (panels (a) and (c)) and for $d=1$, $p_3=1$ (panels (b) and (d)). By repeating these simulations, we obtain $\beta_0 = 0.67 \pm 0.01$ and $\gamma_0 = 2.52 \pm 0.03$ for $d=0$, $p_2=1$ (therefore $D=2)$, and $\beta_1 = 0.52 \pm 0.03$ and $\gamma_1 = 2.85 \pm 0.10$ for $d=1$, $p_3=1$ ($D=3$). In the latter case, due to computational complexity, the simulations were run for a shorter time, as visible in panel (b).}
\label{fig:degre_growth_and_distribution}
\end{figure}

\begin{figure}[h!]
    \centering    
    \subfloat[]{\includegraphics[width=0.44\linewidth]{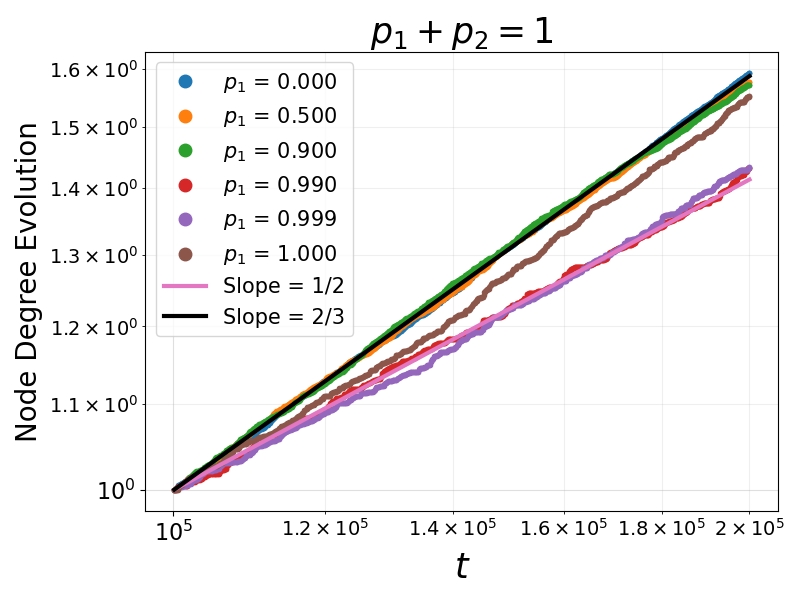}}
    \subfloat[]{\includegraphics[width=0.44\linewidth]{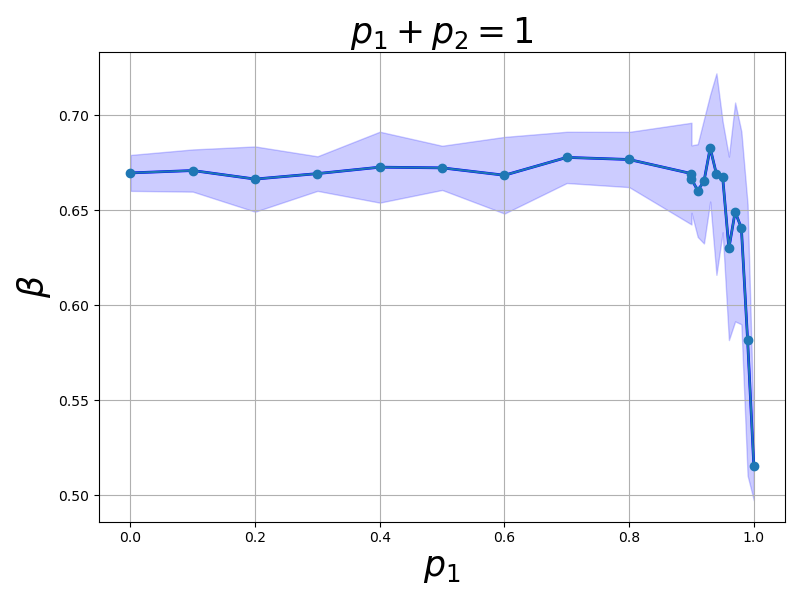}}
    \caption{Panel (a): Degree evolution of the nodes over building time for various values of $p_1$ by keeping $p_1 + p_2 =1$, together with the theoretical slopes $\beta_0=\frac{1}{2}$ for $p_1 = 1$ and $\beta_0=\frac{2}{3}$ for $p_1=0$. As we can see for value $0\leq p_1 \lesssim 0.9$ the growing behavior follows approximately the scaling exponent $2/3$, while for values of $p_1$ extremely close to 1 it follows $\beta_0 = 1/2$. In the intermediate narrow range the the growth is extremely sensitive to the growth realization and approaches $2/3$ as the time increases. Simulation time $T=2\times10^5$.
    Panel (b): Average, over 10 samples, of the value of $\beta_0$ (solid line) together with its error bar, computed as function of $p_1$, being $p_2=1-p_1$. Simulation time $T=10^5$. As the time grows we expect a more pronounced steepening of the numerical transition.}
    \label{fig:degree_evolution_nodes}
\end{figure}

Example of simulations recovering these values of $\beta_d$ and $\gamma_d$ are shown in Fig. \ref{fig:degre_growth_and_distribution}.

Let us emphasize that the computation performed in Appendix~\ref{sec:power_law_exponent_derivation} relies on the condition $p_{D} = 1$ and the remaining $p_{d+1}$ vanish. On the other hand, the degree evolution in the more general setting, where several $p_{d+1}>0$, is instead reported in Appendix \ref{sec:mixed_probabilities_computation}. In the rest of this section we present the result of the numerical simulations in a particular setting, namely when the simplicial complex is built by assuming $p_1, p_2 \in (0,1)$, $p_1+p_2=1$, and we recorded the node degrees, $k_i^{(0)}(t)$, as a function of time growth process; $t$ (note that for $0^\textsuperscript{th}$ $\!\!\!\!$-order simplices the upper-degree correspond to the total degree). The results are reported in Fig.~\ref{fig:degree_evolution_nodes}. We can observe that when $p_1 \to 0$ the value predicted by Eq. \eqref{eq:gamma_power_law} is recovered, indeed in this case we are dealing with a $2$-simplicial complex and thus $\beta_0=D/(D+1)=2/3$. Similarly when $p_1 \to 1$ (Barab\'asi limit), we obtain a $1$-simplicial complex and Eq. \eqref{eq:gamma_power_law} returns $\beta_0=D/(D+1)=1/2$.

For intermediate values of $p_1$ and $p_2$, the theoretical growth of $k_i^{(0)}(t)$ is described by Eq.~\eqref{eq:degree_mixed_solution}. In the asymptotic limit, this leads to a power-law behaviour with exponent $\beta_0 = \frac{2}{3}$ for any $p_2 \neq 0$, while the Barabási--Albert value $\beta_0 = \frac{1}{2}$ is recovered only when $p_2 = 0$.

As shown in Fig.~\ref{fig:degree_evolution_nodes}, this results in an abrupt transition of $\beta_0$ as $p_1 \to 1$. This behaviour originates from the two-term structure of Eq.~\eqref{eq:degree_mixed_solution}: for any finite $p_2$, the contribution scaling as $t^{2/3}$ dominates asymptotically, whereas it disappears when $p_2 = 0$, leaving the standard preferential attachment scaling.
In Sec.~\ref{sec:p1-p2 case} we analytically characterize this abrupt behaviour when the growing of the simplicial complex is modulated by $p_1$ and $p_2$ at the same time.

\subsection{The case: $p_1+p_2=1$}
\label{sec:p1-p2 case}

Let us start from the equation describing the evolution of the node degree when both links and triangles are added to the network.
In compact form we can write
\begin{equation}
\frac{d k_i^{(0)}}{dt} = p_1 \frac{k_i^{(0)}}{\sum_s k_s^{(0)}} + p_2 \sum_{e \ni i} \frac{k_e^{(1)}}{\sum_e k_e^{(1)}}.
\label{eq:deg0_evolution}
\end{equation}
where we denoted by $e \ni i$ the fact that the sum is restricted to links containing node $i$.

Let us now define $T_i$ as the number of triangles incident to node $i$. Since each triangle containing node $i$ contributes to exactly two of its incident edges, we have
\begin{equation}
\sum_{e \ni i} k_e^{(1)} = 2 T_i.
\label{eq:local_sum_edges}
\end{equation}

Similarly, for all the triangles present in the network we have
\begin{equation}
\sum_e k_e^{(1)} = 3 N_2,
\label{eq:global_sum_edges}
\end{equation}
since each triangle contributes to the degree of its three edges.

On average, the number of triangles can be estimated as
\begin{equation}
N_2(t) = p_2 t \quad \implies \quad \sum_e k_e^{(1)} = 3 p_2 t,
\label{eq:N2_growth}
\end{equation}
since at each time step a triangle is added with probability $p_2$.

We can similarly count the number of links in the network. Each time step produces one link with probability $p_1$, and two links when a triangle is added. Therefore
\begin{equation}
N_1(t) = p_1 t + 2 p_2 t = (1+p_2)t.
\label{eq:N1_growth}
\end{equation}

From this, the sum of node degrees is
\begin{equation}
\sum_s k_s^{(0)} = 2 N_1 = 2(1+p_2)t.
\label{eq:sum_deg0}
\end{equation}

We now consider the evolution of the number of triangles incident to node $i$. From Eq.~\eqref{eq:deg0_evolution}, the triangle contribution reads
\begin{equation}
\frac{d T_i}{dt} = p_2 \frac{\sum_{e \ni i} k_e^{(1)}}{\sum_e k_e^{(1)}}.
\end{equation}

By using Eqs.~\eqref{eq:local_sum_edges} and \eqref{eq:N2_growth}, we obtain
\begin{equation}
\frac{d T_i}{dt} = p_2 \frac{2T_i}{3 p_2 t} = \frac{2}{3} \frac{T_i}{t}.
\label{eq:Ti_evolution}
\end{equation}

Eq.~\eqref{eq:Ti_evolution} can be easily solved, yielding
\begin{equation}
T_i(t) = C_i t^{2/3}.
\label{eq:Ti_solution}
\end{equation}

Substituting this result into Eq.~\eqref{eq:deg0_evolution}, we obtain
\begin{equation}
\begin{aligned}
\frac{d k_i^{(0)}}{dt} =& \frac{p_1}{2(1+p_2)} \frac{k_i^{(0)}}{t} + \frac{2}{3} \frac{T_i}{t} \\
=& \frac{p_1}{2(1+p_2)} \frac{k_i^{(0)}}{t} + \frac{2}{3} C_i t^{-1/3}.
\end{aligned}
\label{eq:ki0_intermediate}
\end{equation}

and in compact form, we can write
\begin{equation}
\frac{d k_i^{(0)}}{dt} = a \frac{k_i^{(0)}}{t} + \frac{2}{3} b t^{-1/3},
\label{eq:ki0_final_form}
\end{equation}
whose solution is
\begin{equation}
k_i^{(0)}(t) = A_i t^a + \frac{\frac{2}{3} b}{\frac{2}{3}-a} t^{2/3}.
\label{eq:degree_mixed_solution}
\end{equation}

Since $a \leq \frac{1}{2} < \frac{2}{3}$ (for $p_2>0$), the asymptotic behaviour is dominated by the second term,
\begin{equation}
k_i^{(0)}(t) \sim t^{2/3}.
\label{eq:ki0_asymptotic}
\end{equation}

This result highlights that even a small probability of triangle formation ($0<p_2<<1$) strongly affects the asymptotic growth of node degrees, while if $p_2=0$ then the rightmost term in Eq. \eqref{eq:degree_mixed_solution} disappears and the node degree grows as $t^{a(p_2=0)} = t^{1/2}$. Finally, let us specify that the $\beta_0=2/3$ scaling concerns only the nodes introduced through a triangle-formation event, whereas for the others the second term of Eq. \eqref{eq:degree_mixed_solution} is identically null.

This computation can be further generalized for higher entering dimension and by including the lower degree into the equations (see Appendix~\ref{sec:mixed_probabilities_computation}).

\subsection{Comparison with empirical data}\label{sec:empirical}
\begin{figure}[h!!]
\centering
\subfloat[]{\includegraphics[width=0.45\linewidth]{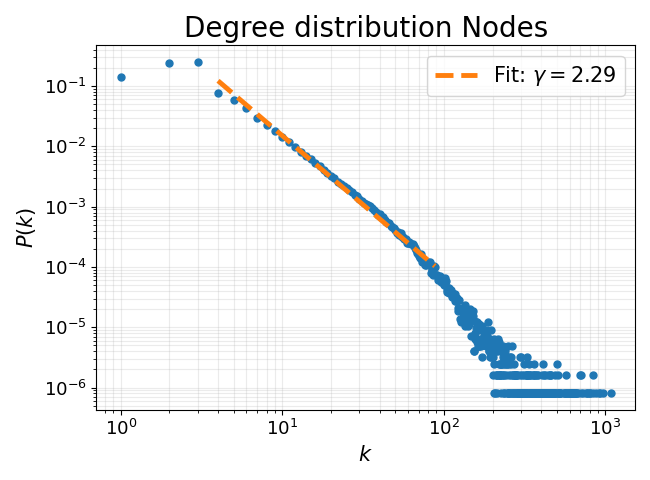}}
\subfloat[]{\includegraphics[width=0.45\linewidth]{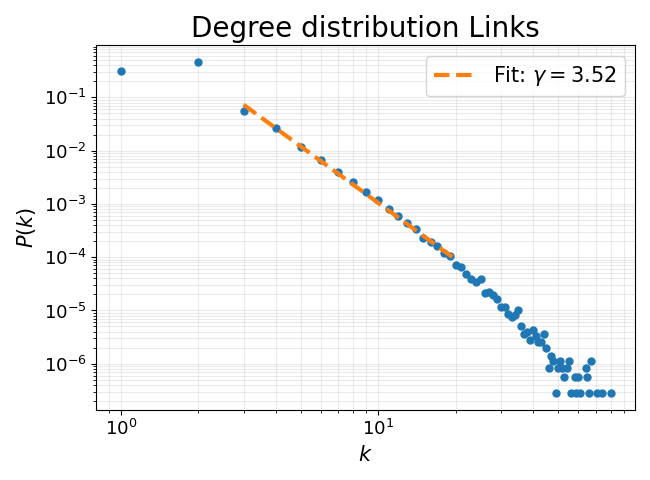}}\\
\subfloat[]
{\includegraphics[width=0.45\linewidth]{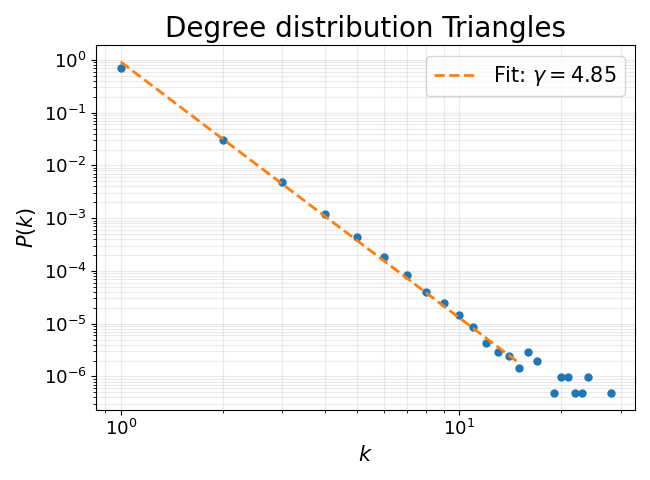}}
    \caption{Fit of the (up-)degree distribution of nodes (panel (a)), links (panel (b)) and triangles (c)) for an empirical simplicial derived from the co-authorship interactions for scientific publication \cite{Benson-2018-simplicial} by fixing the maximal dimension
    $D=3$. The deriving simplicial complex is composed by $N_0=1230962$ nodes, $N_1=3553426$ links, $N_2=2069277$ triangles and 
    $N_3=409313$ tetrahedra. As we can see the empirical power-law exponents are $\gamma_0= 2.29$, $\gamma_1 = 3.52$, $\gamma_2=4.85$ and are qualitative captured by those presented in Eq. \eqref{eq:theoretical_results_for_fit_comparison} also mantaining their corresponding dimension hierarchy.}
\label{fig:empirical_fit}
\end{figure}

We conclude this section by comparing characteristic quantities of the simplicial complexes generated by the algorithm described here with those obtained from empirical data. 
A simple real-world system that can be represented through higher-order interactions is the network of co-authorship collaborations in scientific publications. In this context, papers with multiple authors, represented as nodes, define simplices that can be connected when they share one or more collaborators, finally forming a simplicial complex.
In Fig.~\ref{fig:empirical_fit}, we report the (up-)degree distributions of the nodes, links, and triangles belonging to the empirical simplicial complex constructed from the dataset introduced in \cite{Benson-2018-simplicial}, together with the corresponding power-law fits. (For simplicity, we restrict the analysis to maximum dimension $D=3$.)
The resulting exponents can be compared with those given by Eq.~\eqref{eq:gamma_power_law}, which were analytically derived for \textit{pure} simplicial complexes, i.e., a collection of simplices where every maximal piece shares the exact same dimension (see Appendix~\ref{sec:power_law_exponent_derivation}). However, as shown in Sec.~\ref{sec:p1-p2 case} and generalized in Appendix~\ref{sec:mixed_probabilities_computation_general}, these relations remain valid in the limit of a large number of incoming structures and read for $D=3$:
\begin{equation}
    \begin{aligned}
        \gamma_0 &= 2.\overline{3},\\
        \gamma_1 &= 3,\\
        \gamma_2 &= 5.
    \end{aligned}
    \label{eq:theoretical_results_for_fit_comparison}
\end{equation}
By comparing Eq.~\eqref{eq:theoretical_results_for_fit_comparison} with the empirical fits reported in Fig.~\ref{fig:empirical_fit}, we observe that our theoretical estimates satisfactorily capture the critical exponents obtained from real data. In particular, they closely reproduce the node- and triangle-degree distributions, while also qualitatively describe the link-degree distribution and preserving the hierarchy among the corresponding exponents.
Although the local, "microscopic" interactions among authors do not strictly follow the growth mechanism proposed in this work and illustrated in Fig. \ref{fig:simplex_growth}, the model is still able to capture relevant features of the empirical system.
The residual deviations observed probably originate from finite size corrections and from the underlying generative scheme which can only partially harmonize with the preferential attachment ansatz.
A more comprehensive comparison between the proposed model and empirical datasets is of great interest and is left for future work.

\subsection{Spectral Dimensions}
\label{sec:spectral_dimensions}

\begin{figure}[h!]
    \centering
   \subfloat[]{\includegraphics[width=0.45\linewidth]{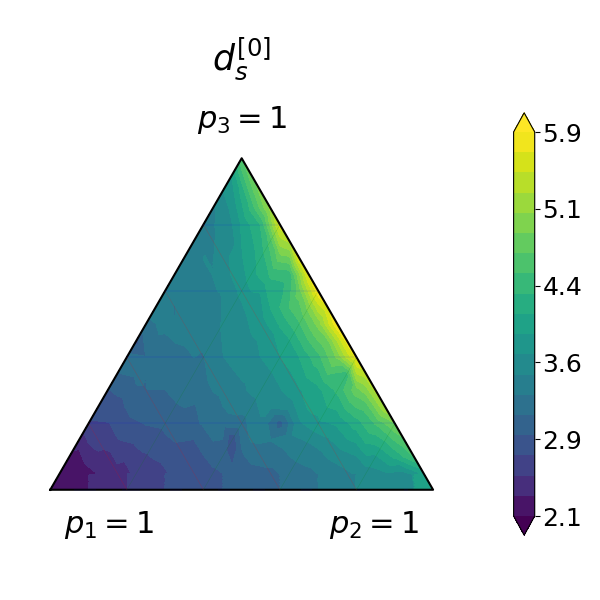}}
    \subfloat[]{\includegraphics[width=0.45\linewidth]{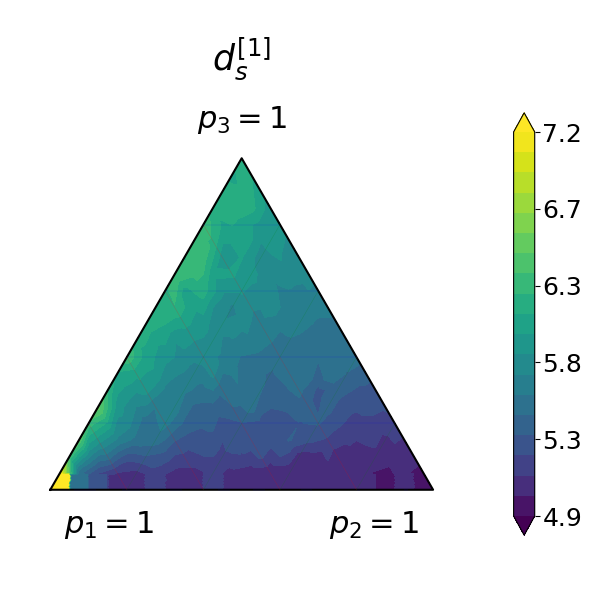}}\\
    \subfloat[]{\includegraphics[width=0.45\linewidth]{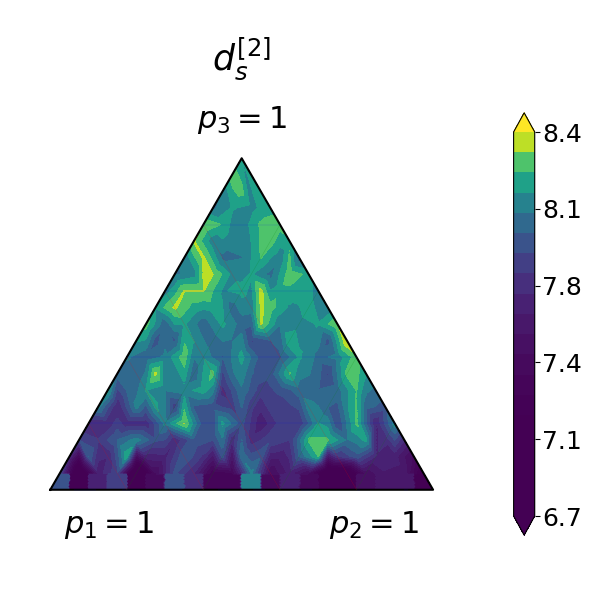}}    
    \subfloat[]{\raisebox{2mm}{\includegraphics[width=0.45\linewidth]{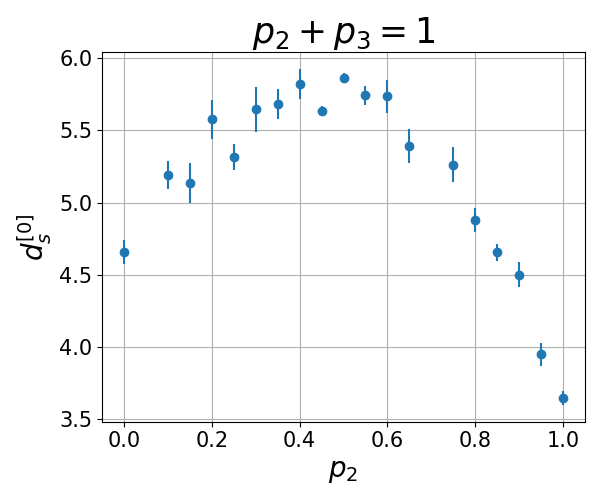}}}

    \caption{Spectral dimensions $d_s^{(0)}$, $d_s^{(1)}$, $d_s^{(2)}$ as a function of the construction parameters $p_1$, $p_2$, $p_3$, with $p_1+p_2+p_3=1$. Note that $d_s^{[1]}$ and $d_s^{[2]}$ are defined, respectively for $p_1=1$ and $p_1+p_2=1$, only on the seed simplicial complex used for the construction.
    In panel (d) is depicted a representative example of non-trivial pattern of the node spectral dimension varying $p_2$ and $p_3$, by keeping $p_2+p_3=1$. (Right-upper side of panel (a)).}
    \label{fig:spectral_dimension}
\end{figure}

It is well known that the topology of the underlying substrate influences the dynamical process running on top of it \cite{Millan2025, febbe2026random}. In this context, we analyze the spectral dimension introduced in Section~\ref{sec:simpcmplx}, that depends on the $d$-Laplace matrix and thus on the boundary operators of the underlying topology.
In Fig.~\ref{fig:spectral_dimension} we report the spectral dimensions $d_s^{(0)}$, $d_s^{(1)}$, $d_s^{(2)}$, as a function of construction parameters $p_1$, $p_2$ $p_3$, with $p_1 + p_2+p_3=1$. Let us note that the results perfectly align with those reported in \cite{torres2020simplicial} in case of $p_3=1$ (see also Tab. \ref{tab:spectral_dimension_flavor}).

\begin{table}[H]
\centering
\begin{tabular}{c|c|}
 & $d_s^{(n)}$ \\
\hline
$n=1$\: &\: 4.88 $\pm$ 0.18\:  \\
\hline
$n=2$ \:& \:6.25 $\pm$ 0.11\: \\
\hline
$n=3$ \:& \:8.1 $\pm$ 0.2\:
\end{tabular}
\caption{Values of the spectral dimension $d_s^{(n)}$ computed for simplicial complexes generated with $p_3=1$, $N_0=2000$, $N_1=5994$ and $N_2=5992$. These are the same parameter setting of \cite{torres2020simplicial} and we obtained results in agreement with those reported there for the higher-order spectral dimensions.}
\label{tab:spectral_dimension_flavor}
\end{table}

\section{Conclusions}\label{sec:conclusion}
In recent years, simplicial complexes have drawn increasing attention in the field of complex systems as mathematical tools capable of describing and encapsulating higher-order interactions.
From a theoretical standpoint, the construction of synthetic simplicial complex structures for modeling studies, data analysis, and benchmarking of algorithms is therefore an important topic in network theory. In particular, we recall the works \cite{bianconi2016network, wu2015emergent, bianconi2017emergent, torres2020simplicial}, where specific classes of simplicial complexes are defined and studied under the so-called preferential attachment regime (among others), and new construction algorithms are introduced for higher-order networks.

Recently, a new operator capable of describing hopping processes across multi-dimensional structures of a simplicial complex has been introduced. By leveraging the topological properties naturally induced by such an operator, we define a new algorithm for the generation of simplicial complexes. By modulating the dimension-related parameters $p_1, \cdots, p_D$, we reproduce a \textit{dimension-wise preferential attachment} mechanism, consistent with existing models in the literature in the appropriate limits, while providing a natural generalization to multiple dimensional orders. As a byproduct, this generative rule can also be viewed as a method for constructing standard dyadic networks that exhibit a power-law degree distribution and regulated clustering coefficient.

The construction algorithm is here extensively described, and the code for simplicial complex generation is freely available in the accompanying repository \cite{diegofebbe_2026_21563618}.

These novel structures have been studied both analytically and numerically in terms of generalized degree growth and asymptotic distributions. We analyze possible transitions arising under different parameter settings and explore specific constructions that may be adapted to particular modeling needs. 
Finally, we investigate spectral dimensions of different orders, quantities that characterize the interplay between dynamical processes and their underlying topology \cite{torres2020simplicial}.

\section*{Acknowledgement}
The work of D. Febbe and D. Fanelli is supported by \#NEXTGENERATIONEU (NGEU) and funded by the Ministry of University and Research (MUR), National Recovery and Resilience Plan (NRRP), project MNESYS (PE0000006) ‘A Multiscale integrated approach to the study of the nervous system in health and disease’ (DR. 1553 11.10.2022).

\section*{Conflict of interest}
The authors declare no competing of interests.

\section*{Data availability statement}
The code of the proposed algorithm used to generate the data for this study is openly available in the repository \cite{diegofebbe_2026_21563618}.

\appendix

\section{Number of paths from $d$ to $D-1$}
\label{sec:number_of_paths}

In this section, we count the number of paths from a $d$-dimensional simplex $\sigma_i^{(d)}$ to a $(D-1)$-dimensional simplex $\sigma_i^{(D-1)}$, which serves as the basis for constructing a new $D$-dimensional simplex with the incoming node, according to the process described in Sec. \ref{sec:topology}. These paths are generated by iteratively applying the adjacency operator $M$ defined in Eq. \eqref{eq:not_normalized_M}, as required by the process described in Eq. \eqref{eq:degree_general_equation}, and are used as a normalization to assess whether the simplex $\sigma_i^{(d)}$ belongs (yielding 1) or does not belong to $\sigma_i^{(D-1)}$ (returning thus 0).
Let us focus on a generic term of Eq. \eqref{eq:degree_general_equation} by considering the product of $D-d-1$ matrices:
\begin{equation}
M_{i,j_{D-d-1}} \dots M_{j_{2}j_{1}}.
\label{eq:product_of_matrices}
\end{equation}
The paths considered here start from the $d$-dimensional simplex and traverse the dimensions until reaching the $(D-1)$-dimensional simplex. First, one can choose among the $(d+1)$-dimensional simplexes contained in the $(D-1)$-simplex, whose number equals the remaining available nodes, $D-d-1$. From each of these, the path can proceed to $(d+2)$-dimensional simplexes, whose number is $D-d-2$, and so on.

Therefore, the total number of paths required to normalize the terms in Eq. \eqref{eq:degree_general_equation} is given by the product of the number of available choices at each step, namely
\begin{equation}
\label{eq:number_of_path_simplicial_complex}
(D-d-1)\cdot (D-d-2)\cdot \dots = (D-d-1)!.
\end{equation}
This normalization can be equivalently implemented by applying an entries-wise characteristic function $\mathbb{1}$ to the matrix product in Eq. \eqref{eq:product_of_matrices}.
\begin{figure}[h!!]
    \centering
    \includegraphics[width=0.9\linewidth]{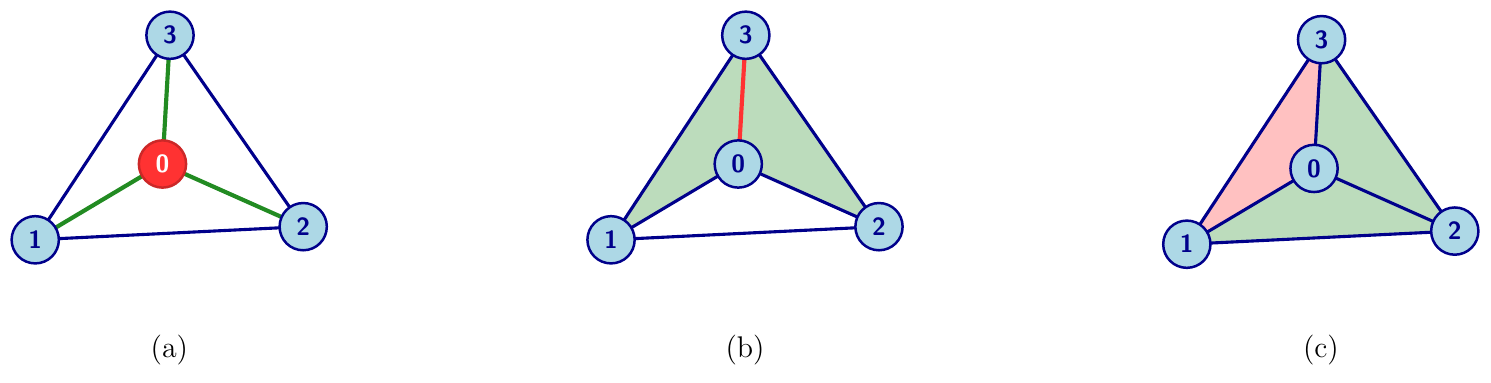}
    \caption{Tetrahedron, basis of the new 4-dimensional structure that will be created. Here, there are 6 paths from node $(0)$ to the tetrahedron $(0,1,2,3)$: from node $(0)$ (red in panel (a)) one can rise to one of the three connected links (green), then from each of these links one can then rise to two connected triangles (panel (b)), and finally to the sole tetrahedron (panel (c)).}
    \label{fig:number_of_paths_example}
\end{figure}

A visual representation for the case $D=4$ is shown in Fig. \ref{fig:number_of_paths_example}. Here, the $3$-dimensional tetrahedron $[0,1,2,3]$ serves as the basis for the construction of a $4$-dimensional simplex by adding a new incoming node $4$ (not shown). Let us assume that we want to study the evolution of the degree of node $0$, so that $\sigma_i^{(d)} = [0]$ with $d = 0$. Its degree increases by one by connecting to the new incoming node $4$, provided that node $0$ belongs to the tetrahedron $[0,1,2,3]$ selected as the basis for the construction (see also Fig. \ref{fig:simplex_growth}).
In general, the question that arises is whether the simplex $\sigma_i^{(d)} = [0]$ belongs or not to the selected basis. Such a connection across distant dimensions can be encoded by the terms appearing in Eq. \eqref{eq:product_of_matrices}, which represent, in this example, that the tetrahedron $[0,1,2,3]$ is connected to triangles, which are connected to links, which in turn are connected to node $0$. However, this product does not return a binary result, but rather the number of paths from the tetrahedron $[0,1,2,3]$ to node $0$.
To estimate the number of such paths, we can proceed as follows: starting from node $0$, we can rise to all the $1$-dimensional structures (links) connected to it, which can be counted by considering all the nodes in $[0,1,2,3]$ except $[0]$ to which they connect forming $1$-dimensional simplices, namely three links ($[0,a]$ for $a$ in $\{1,2,3\}$). From each of these links, we can then rise to higher-dimensional structures (triangles), which can be counted by adding one of the remaining two nodes, and so on.

In general, the number of paths from a $d$-dimensional structure to the $(D-1)$-dimensional construction base ($d=0$ and $D=4$ in the previous example) can be counted as follows. The simplex $\sigma_i^{(d)}$ is connected to all the $(d+1)$-dimensional simplices, whose number amounts to the number of remaining nodes in $\sigma_i^{(D-1)}$, excluding those already belonging to $\sigma_i^{(d)}$, namely $D-d-1$. These $\sigma_i^{(d+1)}$ simplices are in turn connected to the $(d+2)$-dimensional simplices by including one of the remaining $D-d-2$ nodes in the interaction, and so on. By multiplying all the possibilities along the ascending path, we finally obtain Eq. \eqref{eq:number_of_path_simplicial_complex}.

\section{Power law exponent derivation} \label{sec:power_law_exponent_derivation} 
In this section, we derive the growth law for the degree of simplices over time and the power-law exponent of their distribution, as expressed in Eq. \eqref{eq:gamma_power_law}.
Let us recall, as justified in Sec. \ref{sec:topology}, that for simplicity, and in order to avoid introducing dimensional biases, for the growth scheme defined in Fig. \ref{fig:simplex_growth} the degree is chosen to be only the \textit{up}-degree, i.e., $k_d \to k$. This will therefore be the only degree considered throughout this section, since including the \textit{down}-degree in the formulation would result in the addition of dimension-dependent terms in the definition of the simplices selection probabilities $\Pi^{(d)}$.

Let us start by considering some basic examples with low-dimensional values of $d$ and $D$ in Eq. \eqref{eq:degree_general_equation}, and then we will generalize.

\subsection{$d=0$, $p_2=1$} \label{sec:d=0,p2=1}

\begin{figure}[h!]
    \centering
    \includegraphics[width=0.9\linewidth]{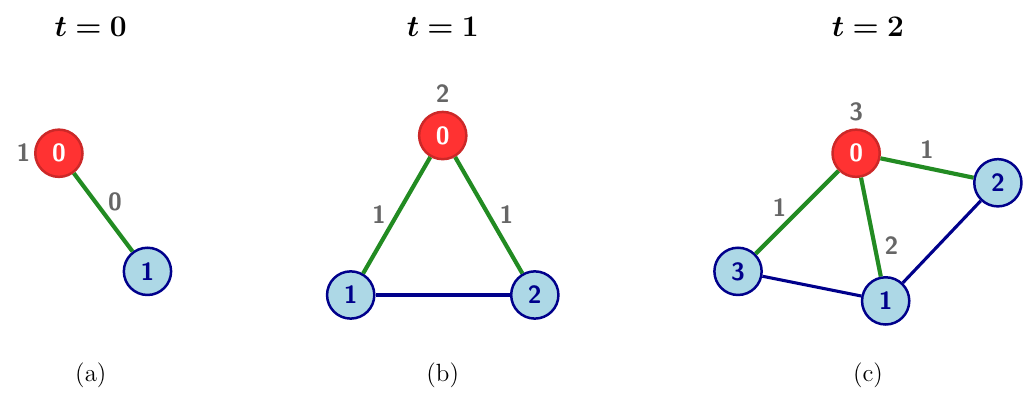}
    \caption{The objective is to understand how to express the sum of the degrees of the links connected (green in the figure) to node 0 (red in the figure) as a function of its degree $k_0^{(0)}$.
    Panel (a): $k_0^{(0)} = 1$ and $\sum_j M_{0j} k^{(1)}_j = 0$, since there are no triangles connected to the link $(0,1)$.
    Panel (b): the first triangle including node 0 is formed by attaching the incoming node to the already present link $(0,1)$. The degree of node 0 increases by 1, so $k_0^{(0)} = 2$, while $\sum_j M_{0j} k^{(1)}_j = 2$.
    Let us note (panel (c)) that for each new triangle that includes node 0, $k_0^{(0)} \to k_0^{(0)} + 1$ and $\sum_j M_{0j} k^{(1)}_j \to \sum_j M_{0j} k^{(1)}_j + 2$, from which Eq. \eqref{eq:degree_links_to_degree_node} can be derived. If instead the triangle formed does not include node 0, neither its degree nor the degree of the links connected to it increases.}
    \label{fig:nodes_to_triangles_degree}
\end{figure}
Here we study the evolution of the node degree ($d = 0$) in cases where the simplicial complex is constructed with the growth parameters set to $p_2 = 1$, $p_i=0$ $\forall$ $i \ne 2$ implying that the resulting simplicial complex has dimension $D = 2$. Note that $p_2 = 1$ also implies that the simplicial complex is formed solely by triangles (and, by the closure property, by the belonging links and nodes).

After these considerations, Eq. \eqref{eq:degree_general_equation} reduces to
\begin{equation}
    \frac{dk_i^{(0)}}{dt} = \sum_j M_{ij} \Pi^{(1)}_j= \frac{\sum_j M_{ij} k^{(1)}_j}{\sum_\ell k_\ell^{(1)}},
    \label{eq:degree_evolution_nodes_triangles}
\end{equation}
which can be solved by expressing the numerator and the denominator as functions of the degree $k_i^{(0)}$ of node $i$.

Let us note, as can be seen from Fig. \ref{fig:nodes_to_triangles_degree}, that when the degree of node $i$ is $k_i^{(0)} = 1$, the sum of the degrees of all the links connected to it is zero, since no triangles involving node $i$ can yet be formed. However, as soon as a new triangle including node $i$ is formed, its degree increases by 1, due to the creation of a new link incident to it, while the sum of the degrees of all the links connected to it increases by 2: one contribution from the newly formed link and one from the already existing link to which the triangle attaches. This process repeats for each new triangle attaching to a link connected to node $i$. See Fig. \ref{fig:nodes_to_triangles_degree} for a visual representation.

From this, we can write
\begin{equation}
    \sum_j M_{ij} k^{(1)}_j = 2(k^{(0)}_i - 1),
    \label{eq:degree_links_to_degree_node}
\end{equation}
and by summing over all the nodes,
\begin{equation}
    \sum_\ell k_\ell^{(1)} = \frac{\sum_\ell2(k^{(0)}_\ell - 1)}{2}=\sum_\ell(k^{(0)}_\ell - 1),
    \label{eq:normalization_degree_links_to_nodes}
\end{equation}
where the division by 2 accounts for the multiple counting of the link degrees by their two endpoint nodes.

By substituting Eqs. \eqref{eq:degree_links_to_degree_node} and \eqref{eq:normalization_degree_links_to_nodes} into Eq. \eqref{eq:degree_evolution_nodes_triangles}, we obtain:
\begin{equation}
    \frac{dk_i^{(0)}}{dt} = \frac{2(k^{(0)}_i - 1)}{\sum_j (k^{(0)}_j - 1)} = \frac{2(k^{(0)}_i - 1)}{4t-t} \sim \frac{2k^{(0)}}{3t}.
    \label{eq:final_eq_nodes_triangles}
\end{equation}
Since at each time step a new node is added, we have $\sum_{\text{nodes}} 1 = t$, while the total node degree increases by 4: 2 for the new node, and 1 for each node belonging to the link to which the new triangle is attached.

The solution of Eq. \eqref{eq:final_eq_nodes_triangles} is
\begin{equation}
k^{(0)}_i (t) = k^{(0)}_i (t_i)\left(\frac{t}{t_i}\right)^\beta\!\!,
\label{eq:final_eq_nodes_triangles_solution}
\end{equation}
where $t_i$ denotes the entrance time of node $i$, $k^{(0)}_i (t_i)=2$ and $\beta = \frac{2}{3}$.
Since the node degree given by Eq. \eqref{eq:final_eq_nodes_triangles_solution} increases monotonically, the number of nodes having degree greater than a given $k_i^{(0)}$ is determined by $t_i = T \left(\frac{2}{k_i}\right)^{1/\beta},$ as at each time step a new node enters the simplicial complex, with $T$ indicating the total construction time. Therefore, the probability of picking a node with degree lower than $k_i$ is 
\begin{equation}
    P(k) = 1-\left(\frac{2}{k_i}\right)^{1/\beta}\!\!\!\!\!\!,
    \label{eq:prob_node_degree_lower}
\end{equation}
and by differentiating Eq. \eqref{eq:prob_node_degree_lower}, we obtain the degree probability distribution (see \cite{barabasi2016network})
\begin{equation}
    p_k \sim k^{-\left(1+\frac{1}{\beta}\right)} \equiv k^{-\gamma}.
\end{equation}

\subsubsection{Discrete time process}\label{sec:appendix_discrete_version}
The aim of this section is to show that an equivalent result can be obtained by considering time to be a discrete variable instead of a continuous one as done above. We can thus write the discrete version of Eq.~\eqref{eq:degree_evolution_nodes_triangles} with a generic $\beta$ as follows:
\begin{equation}
k_i(t+1) = k_i(t) + \beta \frac{k_i(t)}{t} = k_i(t)\left(1+\frac{\beta}{t}\right).
\end{equation}
By recursion, we can write $k_i(t+1)$ as a function of $k_i(\tau=1)$:
\begin{equation}
k_i(t+1) = k_i(\tau=1)\prod_{\tau=1}^{t} \left(1+\frac{\beta}{\tau}\right)
= k_i(\tau=1)\prod_{\tau=1}^{t} \left(\frac{\tau+\beta}{\tau}\right).
\end{equation}
We can now rewrite the productorial term as
\begin{equation}
\prod_{\tau=1}^{t} \left(\frac{\tau+\beta}{\tau}\right) = \frac{\Gamma(t+\beta+1)}{\Gamma(\beta+1)\Gamma(t+1)},
\end{equation}
where we have made use of the Euler $\Gamma$ function, whose asymptotic behavior can be estimated through Stirling’s formula:
\begin{equation}
\frac{\Gamma(t+\beta+1)}{\Gamma(\beta+1)\Gamma(t+1)}
\sim \frac{\sqrt{2\pi (t+\beta)} \left(\frac{t+\beta}{e}\right)^{t+\beta}}{\sqrt{2\pi t} \left(\frac{t}{e}\right)^{t}}
\sim t^{\beta}.
\end{equation}
Thus, we recover the same growth regime for large values of the building time as in Eq. \eqref{eq:final_eq_nodes_triangles_solution}.

\subsection{$d=0$, $p_3=1$} \label{sec:d=0,p3=1}

\begin{figure}[h!!]
    \centering
    \includegraphics[width=0.9\linewidth]{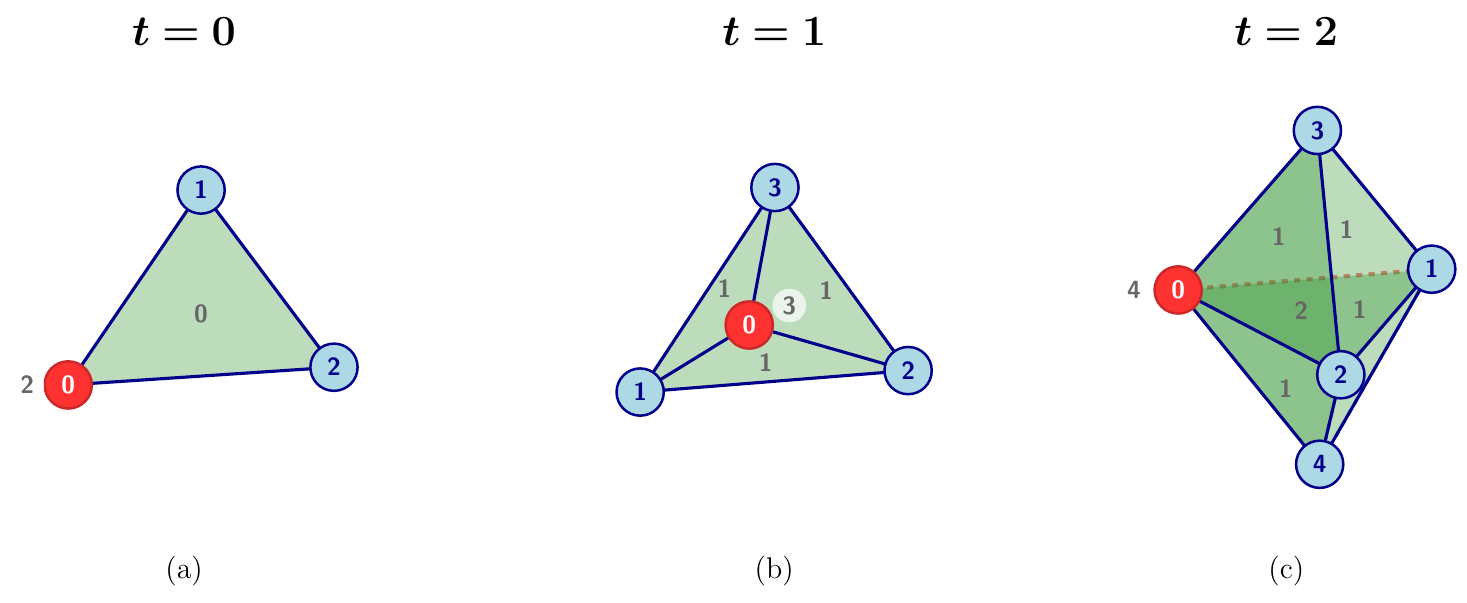}
    \caption{Here we want to write the sum of the degrees of all triangles that include node $0$, namely $\sum_{h,j} \mathbb{1} ( M_{0j} M_{jh}) k_{h}^{(2)}$, as a function of its degree $k_0^{(0)}$. In this way, we can express the sum of the degrees of the triangles (shown in green) incident to a given node $i$ (the red node labelled as 0 in the figure) in terms of its degree, thereby obtaining a closed solution for Eq. \eqref{eq:nodes_to_triangles_degree_}. The degree of node 0 is reported next to it (2 in panel (a), 3 in panel (b), and 4 in panel (c)), while the (up-)degree of each incident triangle is indicated by the numbers 0, 1, and 2 displayed on the corresponding green filled triangular regions.} In panel (a), we have the seed simplex [0,1,2] and no tetrahedra have yet been formed, so $k_0^{(0)} = 2$ (red in the figure), and the only triangle formed has degree 0 (green in the figure). In panel (b), with the new incoming node (labelled as 3), a tetrahedron containing node $0$ is formed. Thus, $k_0^{(0)} \to k_0^{(0)} + 1 = 3$, while $\sum_{h,j} \mathbb{1} ( M_{0j} M_{jh}) k_{h}^{(2)} = 3$, one for each of the three triangles containing node $0$.
    A strategy to count such structures will be important for the general case. Here, we note that triangles containing node $0$ can be formed by considering it together with a pair of the remaining nodes as vertices. These rules for the degree evolution can be generalized for new incoming tetrahedra that include node $0$ (panel (c)), giving $k_0^{(0)} \to k_0^{(0)} + 1$ and $\sum_{h,j} \mathbb{1} ( M_{0j} M_{jh}) k_{h}^{(2)} \to \sum_{h,j} \mathbb{1} ( M_{0j} M_{jh}) k_{h}^{(2)} + 3$. Indeed, the structure in panel (c) can be seen as two tetrahedra, as shown in panel (b), that share the triangle $(0,1,2)$, whose degree is therefore 2.
    \label{fig:nodes_to_tetra_degree}
\end{figure}

Here we want to study the evolution of the degree of the nodes ($d=0$) in the case $p_3 = 1$, meaning that the resulting simplicial complexes will be formed solely by tetrahedra (and all the sub-simplices required by the closure property). In order to solve Eq. \eqref{eq:degree_general_equation} for this choice of parameters, it is important to express the sum of the degrees of the triangles containing node $i$ (node labelled 0 in Fig. \ref{fig:nodes_to_tetra_degree}) as a function of its degree . These triangles may serve as bases for the construction of tetrahedra.

Following the computation in Sec. \ref{sec:d=0,p2=1} (see also Fig. \ref{fig:nodes_to_tetra_degree}), we can start from the trivial case where the degree $k_i^{(0)}$ of node $i$ is equal to 2 (panel (a) in Fig. \ref{fig:nodes_to_tetra_degree}). The total degree of the triangles containing it is therefore 0, as there are no tetrahedra that can be formed with node $i$ ($t=0$). As a new tetrahedron attaches to a triangle containing node $i$, the degree of the connected triangles increases by 3 ($t=1$, panel (b) in Fig. \ref{fig:nodes_to_tetra_degree}). Indeed, all the triangles formed by taking node $i$ together with a pair of the remaining nodes are connected once to the tetrahedron, and, as before, this process repeats for all incoming tetrahedra ($t=2$, panel (c) in Fig. \ref{fig:nodes_to_tetra_degree}) that include node $i$.

From this, we can write
\begin{equation}
    \frac{dk_i^{(0)}}{dt} = \sum_{h,j} \mathbb{1} ( M_{ij} M_{jh})\Pi_{h}^{(2)} = \frac{3(k^{(0)}_i - 2)}{\frac{\sum_j 3(k^{(0)}_j - 2)}{3} } = \frac{3(k^{(0)}_i - 2)}{6t -2t} \sim\frac{3k^{(0)}_i}{4t},
    \label{eq:nodes_to_triangles_degree_}
\end{equation}
where, in this case we divide by 3 because each triangle is counted three times (once for each node), and the node degree increases by 6 at each time step: 3 for the new incoming node connecting to a triangle, and 1 for each of the already present vertices.
\subsection{$d=1$, $p_3=1$} \label{sec:d=1,p3=1}

\begin{figure} [h!]
    \centering
    \includegraphics[width=0.9\linewidth]{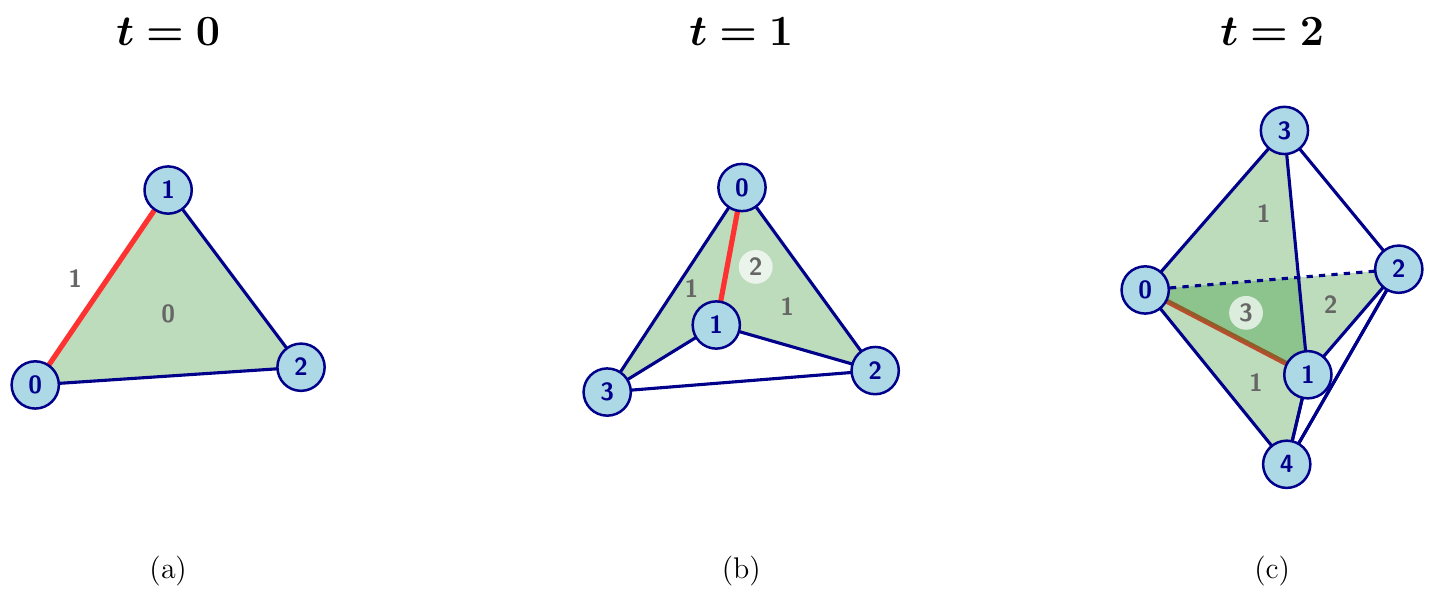}
    \caption{Here we want to express how the degree $k_{0}^{(1)}$ of the link $(0,1)$ (with index $i=0$, red in the figure) regulates the sum of the degrees of the triangles connected to it, $\sum_j M_{0j} k^{(2)}_j$ (green in the figure), in the growing process defined by the parameter $p_3 = 1$. In the trivial case (panel (a)), before the first tetrahedron is formed, $k_{0}^{(1)} = 1$, since link $(0,1)$ is connected to the triangle $(0,1,2)$, while $\sum_j M_{0j} k^{(2)}_j = 0$, as there are no tetrahedra available to connect with. Note that if, as an initial condition, link $(0,1)$ does not belong to any triangle, its degree will remain constant.
    In panel (b), the first tetrahedron attached to triangle $(0,1,2)$ is formed, so $k_{0}^{(1)} \to k_{0}^{(1)} + 1$, as it forms a new triangle with the incoming node, while $\sum_j M_{0j} k^{(2)}_j \to \sum_j M_{0j} k^{(2)}_j + 2$. As in the previous cases, for newly forming tetrahedra (panel (c)) the same pattern repeats, as reported in Eq. \eqref{eq:links_evolution_computation}.}
    \label{fig:links_to_tetra_degree}
\end{figure}
Let us study the growth law for the degree of a link $i$ ($d=1$) when $p_3 = 1$.
As in the previous cases, we start from the trivial structure and build from there. We can note (see Fig. \ref{fig:links_to_tetra_degree}) that when the degree of link $i$ is $k^{(1)}_i = 1$, the triangle attached to it must have degree 0, as there are no tetrahedra containing link $i$. For any new forming tetrahedron that contains link $i$, its degree increases by 1, while the degree of the triangles attached to it increases by 2: one for the new triangle and one for the already existing one.

\begin{figure}[h!]
    \centering
    \includegraphics[width=0.4\linewidth]{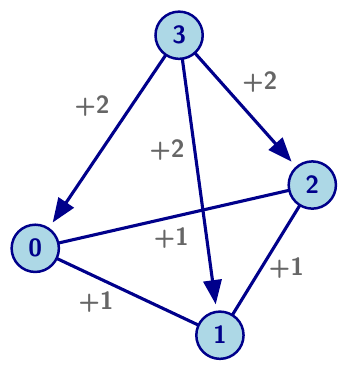}
    \caption{When a new tetrahedron is formed, three new links (dashed in the figure) appear. Their degree is 2, as each is connected to two triangles. Meanwhile, each of the links belonging to the triangle basis connects to a new triangle.}
    \label{fig:degree_links_increase_one_time_step}
\end{figure}

Thus, we obtain
\begin{equation}
    \frac{dk_i^{(1)}}{dt} = \sum_j M_{ij} \Pi^{(2)}_j = \frac{2(k^{(1)}_i - 1)}{\frac{\sum_j 2(k^{(1)}_j - 1)}{3}} = \frac{3(k^{(1)}_i - 1)}{9t-3t} \sim \frac{k^{(1)}}{2t}.
    \label{eq:links_evolution_computation}
\end{equation}
Here, when we write $\sum_\ell k_\ell^{(2)}$ in terms of the link degrees, we are counting each triangle three times, once for each link (so we should divide by 3). Furthermore, note that the total link degree increases by 9 at each time step: 1 for each already present link in the triangle basis, and 2 for each new link of the incoming tetrahedron. Additionally, at each time step 3 new links are formed, so $\sum_{\text{links}} 1 = 3t$ (see Fig. \ref{fig:degree_links_increase_one_time_step}).

\subsection{General case}
\label{sec:general_case_demonstration}
From Secs. \ref{sec:d=0,p2=1}, \ref{sec:d=0,p3=1}, and \ref{sec:d=1,p3=1}, we now derive a general formula describing the evolution of structures of dimension $d$ in cases in which simplicial complexes of dimension $D$ are built by setting $p_D=1$ in Eq. \eqref{eq:degree_general_equation}, with $D>d$.

By summarizing the methods previously presented, in order to write a solvable equation for the degree $k_i^{(d)}$, it is crucial to express the sum of the degrees of all the $(D-1)$-dimensional structures as a function of it. These $(D-1)$-dimensional structures indeed serve as bases for the incoming simplex, and they are selected by a preferential attachment mechanism.

Let us consider the general case, keeping in mind, for instance, Sec.~\ref{sec:d=0,p3=1}, where we expressed the sum of the degrees of the triangles, which serve as bases for tetrahedra, connected to node $i$ as a function of $k_i^{(0)}$. This was done by first considering the initial condition, corresponding to the degree $k_i^{(0)}$ before the first tetrahedron arrives, and then by counting the number of triangles connected to $i$ in a newly formed $3$-simplex. This procedure is repeated for each new tetrahedron attaching to triangles that contain node $i$. Here, we generalize this approach.

As before, let us note that when the first $D$-dimensional structure has not yet been formed, the degree of the $d$-dimensional structure is given by the number of nodes with which the $d$-simplex, having $d+1$ nodes, can connect within the $(D-1)$-dimensional simplex, namely $k_i^{(d)} = (D-1)+1-(d+1)=D-d-1$.

Now, for each new $D$-dimensional structure $\sigma^{(D)}$ containing the simplex $\sigma_i^{(d)}$, the total degree of the $(D-1)$-dimensional structures, which are bases of $\sigma^{(D)}$ and contain $\sigma_i^{(d)}$, can be counted by considering all the remaining nodes of $\sigma^{(D)}$ that are not already included in $\sigma_i^{(d)}$. By excluding these nodes one by one and considering the remaining ones, we can construct and count the $(D-1)$-dimensional structures with degree 1, whose number is $D-d$. In Sec. \ref{sec:d=0,p3=1}, for instance, we counted the three triangles connected to node $i$ as the total number of nodes in a tetrahedron, i.e. 4, minus 1 for node $i$. By excluding one of these three nodes at a time, we obtained the three possible combinations of vertex pairs that, together with node $i$, form the counted triangles.

So, we can write
\begin{equation}
\frac{dk^{(d)}_i}{dt} = \frac{\sum_{j_1, \dots,j_{D-d-1}} \mathbb{1}(M_{i,j_{D-d-1}} \dots M_{j_{2}j_{1}})k_{j_1}^{(D-1)}}{\sum_jk_j^{(D-1)}} = \frac{(D-d)(k_i^{(d)}-D+d+1)}{\frac{\sum_j (D-d)(k_j^{(d)}-D+d+1)}{\binom{D}{d+1}}},
\label{eq:degree_evolution_general_p_D=1}
\end{equation}
where the term $\binom{D}{d+1}$ accounts for the number of times the degree of the structures $\sigma_j^{(D-1)}$, having $D$ nodes, is counted by considering the degree of all the sub-$d$-dimensional structures, that have $d+1$ nodes.
We can summarize the various steps leading to the terms of Eq. \eqref{eq:degree_evolution_general_p_D=1} as:
\begin{itemize}
\item We want to express the sum of the degrees of all the $(D-1)$-dimensional structures, to which the incoming $D$-simplex attaches at each time step, as a function of the degree $k_i^{(d)}$ of $\sigma_i^{(d)}$ belonging to them. So, for the numerator, we can note that (see the examples reported in Figs. \ref{fig:nodes_to_triangles_degree}, \ref{fig:nodes_to_tetra_degree} and \ref{fig:links_to_tetra_degree}):
\begin{itemize}
\item if the sum of those $k^{(D-1)}$ is equal to zero, then $k_i^{(d)}=D-d-1$;
\item in a $D$-simplex the number of $\sigma^{(D-1)}$ simplices to which $\sigma_i^{(d)}$ is contained is equal to $D-d$.
\end{itemize}
\item In the denominator, instead, we count the degree of each $k_j^{(D-1)}$ structure by iteratively summing the numerator for each $\sigma_j^{(d)}$ structure contained in it, namely $\binom{D}{d+1}$ structures.
\end{itemize}

As the last steps, we must evaluate the terms $\sum_{\sigma_j^{(d)}} k_j^{(d)}$ and $\sum_{\sigma_j^{(d)}} 1$ present at the denominator, varying as the growing process evolves, namely the degree and the number of the $d$-dimensional simplices.
\paragraph{}

The former term can be expressed as
\begin{equation}
\label{eq:increase_degree_general_demonstration}
\left(\sum_{\sigma_j^{(d)}} k_j^{(d)}(t+1)-\sum_{\sigma_j^{(d)}} k_j^{(d)}(t)\right)t,
\end{equation}
namely, the increase of the degrees of all simplices of dimension $d$ over time. At each time-step, the term in parentheses in Eq. \eqref{eq:increase_degree_general_demonstration}, constant over time since $p_D=1$, is given by the sum of the degrees of each $d$-dimensional structure within $\sigma^{(D)}$, minus the total degree already present on the $D-1$-dimensional attaching face:
\begin{equation}
 \binom{D+1}{d+1} (D+1-(d+1)) - \binom{D}{d+1} (D-d-1) = (d+2) \binom{D}{d+1},
\label{eq:increase_degree_dimonstration}
\end{equation}
where the first product of each terms accounts for the number of $d$-dimensional simplices, while the second for their degree.

\paragraph{}
Finally, we can count $\sum_{\sigma_j^{(d)}} 1$ as the number of $d$ simplices in a $D$-dimensional structure minus that already present on a $D-1$-dimensional one, namely:
\begin{equation}
\sum_{\sigma_j^{(d)}} 1 = \binom{D+1}{d+1} - \binom{D}{d+1} = \binom{D}{d}.
\end{equation}
By putting everything together in Eq. \eqref{eq:degree_evolution_general_p_D=1}, we obtain:
\begin{equation}
\beta = \frac{\binom{D}{d+1}}{\left(\binom{D+1}{d+1}(D-d)-\binom{D}{d+1}(D-d-1)\right)-(D-d-1)\binom{D}{d}}= \frac{D-d}{D+1},
\label{eq:beta_appendix}
\end{equation}

and therefore:
\begin{equation}
\gamma = 1 + \frac{D+1}{D-d}.
\label{eq:gamm_appendix}
\end{equation}

Numerical simulations for some example cases can be seen in Fig. \ref{fig:degre_growth_and_distribution}.

Finally, we observe that the proposed computation is fully consistent with the generation algorithm illustrated in \cite{bianconi2016network}, where the degree of a simplex $\sigma_i^{(d)}$, denoted here for simplicity by $k_{BR_i}$ (Bianconi--Rahmede), is counted as the number of $D$-simplexes incident to it.
Here, instead, since the topology is induced by operator used to define the Random Walk process defined in \cite{febbe2026random}, which allows the hopper to jump up and down by only one dimension, the degree $k_{FFC_i}$ of the simplex $\sigma_i^{(d)}$ is given by the number of $(d+1)$-dimensional simplices incident to it.

In a single $D$-dimensional simplex, where $k_{BR_i}=1$, the simplex $\sigma_i^{(d)}$ is connected, in our framework, to all the $(d+1)$-dimensional structures containing it, whose number is $D-d$.
Since this holds for each incident $D$-dimensional simplex, we can write

\begin{equation}
\begin{aligned}
&k_{FFC} - k_{BR} = D-d-1,\\
&k_{BR} = k_{FFC} - (D-d-1).
\end{aligned}
\end{equation}

By substituting this relation into Eq. \eqref{eq:degree_evolution_general_p_D=1}, we obtain:
\begin{equation}
\frac{dk^{(d)}_{BR_i}}{dt} = \binom{D}{d+1}\frac{k_{BR_i}^{(d)}}{\sum_j k_{BR_j}^{(d)}},
\label{eq:degree_evolution_bianconi}
\end{equation}
where here,
\begin{equation}
\sum_j k_{BR_j}^{(d)} = \binom{D+1}{d+1}t,
\end{equation}
since each incoming $D$-dimensional simplex increases by 1 the degree of each of its $\binom{D+1}{d+1}$ incident $d$-dimensional simplices.

We can then conclude with the expansion
\begin{equation}
\beta = \frac{\binom{D}{d+1}}{\binom{D+1}{d+1}}=\frac{D-d}{D+1},
\end{equation}
which yields Eq. \eqref{eq:beta_appendix}.

\section{Mixed probabilities computation}
\label{sec:mixed_probabilities_computation}
In Sec.~\ref{sec:p1-p2 case} we analyzed the behaviour of the growth exponent $\beta_0$ when the simplicial complex construction is modulated by the two parameters $p_1$ and $p_2$ at the same time. Here we extend the analysis to more general settings.

In Sec.~\ref{sec:mixed_probabilities_computation_general} we further extend the computation of the growth exponent presented in Sec.~\ref{sec:p1-p2 case} to the case of higher-order construction probabilities.

In Appendix~\ref{sec:Lower degree included} we include the lower-degree term modulating the attachment mechanism for the computation exposed in Sec.~\ref{sec:p1-p2 case}. Although, from a conceptual perspective, this may introduce a dimension-wise bias, since higher-order simplices have a larger lower degree, this modification resolves the abrupt transition shown in Fig.~\ref{fig:simplex_growth}.

Before continuing, notice that we can slightly modify the mechanism described in Eq. \eqref{eq:deg0_evolution}, by introducing a parameter $m$ that controls the number of connections established by each newly added node (as in the celebrated Barab\'asi--Albert model~\cite{barabasi2016network}), we obtain
\begin{equation}
\frac{d k_i^{(0)}}{dt} = m p_1 \frac{k_i^{(0)}}{\sum_s k_s^{(0)}} + m p_2 \sum_{e \ni i} \frac{k_e^{(1)}}{\sum_e k_e^{(1)}}.
\label{eq:deg0_evolution_m}
\end{equation}

However, since
\begin{equation}
\begin{aligned}
N_1(t) &= m p_1 t + 2 m p_2 t, \\
N_2(t) &= m p_2 t,
\end{aligned}
\end{equation}
the factor $m$ cancels out in the normalization terms, leading to the same asymptotic growth behaviour as in the case $m=1$.

Interestingly, under this setting, configurations may arise in which the three links forming a triangle are present without the corresponding 2-dimensional face.
\subsection{General case}
\label{sec:mixed_probabilities_computation_general}

We now consider the general case in which simplices of different dimensions can enter the network. The evolution of the node degree can be written as
\begin{equation}
\frac{d k_i^{(0)}}{dt} = \sum_{d=1}^D p_d \frac{\sum_{\sigma^{(d-1)} \ni i} k_{\sigma}^{(d-1)}}{\sum_{\sigma^{(d-1)}} k_{\sigma}^{(d-1)}}.
\label{eq:general_start}
\end{equation}

where we used the notation $\sigma^{(d-1)} \ni i$ to denote the fact that the sum is restricted to $(d-1)$-simplexes containing node $i$.
As in the previous case, it is useful to introduce the number of simplices $S_i^{(d)}$ containing node $i$. Then we have the identity (see Appendix~\ref{sec:general_case_demonstration})
\begin{equation}
\sum_{\sigma^{(d-1)} \ni i} k_{\sigma}^{(d-1)} = d S_i^{(d)}.
\label{eq:relation_kS}
\end{equation}

Moreover,
\begin{equation}
\sum_{\sigma^{(d-1)}} k_{\sigma}^{(d-1)} = (d+1) N_d.
\label{eq:sum_general}
\end{equation}

The evolution of $S_i^{(q)}$ can therefore be written as
\begin{equation}
\frac{d S_i^{(q)}}{dt} = \sum_{d=q}^D p_d \binom{d-1}{q-1}\frac{\sum_{\sigma^{(d-1)} \ni i} k_{\sigma}^{(d-1)}}{\sum_{\sigma^{(d-1)}} k_{\sigma}^{(d-1)}},
\label{eq:Si_general_original}
\end{equation}

and by defining
\begin{equation}
\alpha_d = \sum_{m=d}^D p_m \binom{m}{d},
\label{eq:alpha_def}
\end{equation}

we can estimate the average number of $r$-simplices as
\begin{equation}
N_d(t) = \alpha_d t.
\label{eq:Nr_growth}
\end{equation}

Substituting Eqs.~\eqref{eq:relation_kS}, \eqref{eq:sum_general}, and \eqref{eq:Nr_growth} into Eq.~\eqref{eq:Si_general_original}, we obtain
\begin{equation}
\frac{d S_i^{(q)}}{dt} = \sum_{d=q}^D \frac{p_d d}{(d+1)\alpha_d} \binom{d-1}{q-1} \frac{S_i^{(d)}}{t}.
\label{eq:Si_general}
\end{equation}

and therefore
\begin{equation}
\frac{d k_i^{(0)}}{dt} = \sum_{d=1}^D \frac{p_d d}{(d+1)\alpha_d} \frac{S_i^{(d)}}{t}.
\label{eq:k_expressed_with_S}
\end{equation}

By solving the triangular system of differential equations for $S_i^{(q)}$, we can determine the degree evolution $k_i^{(0)}$.

In the case $q = D$, Eq.~\eqref{eq:Si_general} reduces to
\begin{equation}
\frac{d S_i^{(D)}}{dt} = \frac{D}{D+1} \frac{S_i^{(D)}}{t}
\implies S_i^{(D)} \sim t^{D/(D+1)},
\label{eq:case_q=D_S_i}
\end{equation}

while, more generally,
\begin{equation}
\dot{\mathbf{S}}_i = \frac{1}{t} A \mathbf{S}_i,
\label{eq:matrix_general}
\end{equation}

with upper triangular matrix with components
\begin{equation}
A_{qd} = \frac{p_d d}{(d+1)\alpha_d}\binom{d-1}{q-1}.
\label{eq:triangular_matrix_component}
\end{equation}

The asymptotic behaviour is determined by the largest eigenvalue of $A$, given by its diagonal entries
\begin{equation}
\lambda_q = \frac{p_q q}{(q+1)\alpha_q}.
\label{eq:diag_entries}
\end{equation}

For $q=D$, $\lambda_D=\frac{D}{D+1}$, while for $q<D$ we have $\alpha_q \ge p_q$ (see Eq. \eqref{eq:alpha_def}), which implies
\begin{equation}
\lambda_q \leq \frac{q}{q+1} < \frac{D}{D+1}.
\label{eq:bound_diag}
\end{equation}

Hence, the largest eigenvalue corresponds to $q=D$, and the highest-dimensional simplices entering the network determine the asymptotic growth of the degree of nodes contained in the $D$-simplices.

\subsection{Lower degree included}
\label{sec:Lower degree included}

The two terms growth mechanisms derived from Eq. \eqref{eq:degree_mixed_solution} only applies to nodes that already belong to at least one triangle. If a node $i$ is introduced at time $t_{i_0}$ without being part of any triangle, then $\sum_{e \ni i} k_e^{(1)} = 0$ for all $t>t_{i_0}$, and the triangle-driven growth channel is effectively suppressed.

In the following, we consider a modified model in which also links not belonging to triangles can be selected.

A natural choice is to include both upper and lower degree contributions, i.e. $K = k_{up}+k_{down}$ and the emerging terms of the equations are of the same type as those present in model \cite{price1965networks}. In the course of this section we will restore the notation $k_i^{(d)} \to k_{{up}_i}^{(d)}$ to indicate the upper degree of a simplex $\sigma_i^{(d)}$ in order to be clearly differentiated from the down degree $k_{{down}_i}^{(d)}$.

With the choice $K = k_{up}+k_{down}$, and $k_{down}=2$, we can write the general node degree evolution as
\begin{equation}
\frac{d k_{{up}_i}^{(0)}}{dt} = p_1 \frac{k_{{up}_i}^{(0)}}{\sum_j k_{{up}_j}^{(0)}} + p_2 \sum_{e \ni i} \frac{k_{{up}_e}^{(1)} + 2}{\sum_e (k_{{up}_e}^{(1)}+2)}.
\label{eq:deg0_shifted}
\end{equation}

As in the previous section, we can compute the number of triangles connected to node $i$:
\begin{equation}
\sum_{e \ni i} k_{{up}_e}^{(1)} = 2T_i.
\label{eq:local_sum_edges2}
\end{equation}

We can also evaluate the numerator of the second term appearing in Eq.~\eqref{eq:deg0_shifted}:
\begin{equation}
\sum_{e \ni i} (k_{{up}_e}^{(1)}+2) = 2T_i + 2k_{{up}_i}^{(0)},
\label{eq:local_shifted}
\end{equation}
where the sum runs over the $k_{{up}_i}^{(0)}$ edges incident to node $i$.

For the denominator instead, by using Eq. \eqref{eq:global_sum_edges}, we have
\begin{equation}
\sum_e (k_{{up}_e}^{(1)}+2) = \sum_e k_{{up}_e}^{(1)} + 2N_1 = 3N_2 + 2N_1,
\label{eq:global_shifted}
\end{equation}

and by substituting Eqs.~\eqref{eq:N2_growth} and \eqref{eq:N1_growth}, we obtain
\begin{equation}
\sum_e (k_{{up}_e}^{(1)}+2) = 3p_2 t + 2(1+p_2)t = (2+5p_2)t.
\label{eq:denominator_shifted}
\end{equation}

By combining all contributions in Eq.~\eqref{eq:deg0_shifted}, we obtain
\begin{equation}
\frac{dk_{{up}_i}^{(0)}}{dt} = \frac{p_1}{2(1+p_2)} \frac{k_{{up}_i}^{(0)}}{t} + p_2 \frac{2T_i + 2k_{{up}_i}^{(0)}}{(2+5p_2)t}.
\label{eq:ki0_shifted_final}
\end{equation}

Similarly, the growth of triangles connected to $i$ is governed by
\begin{equation}
\frac{d T_i}{dt} = p_2 \frac{2T_i + 2k_{{up}_i}^{(0)}}{(2+5p_2)t}.
\label{eq:Ti_shifted}
\end{equation}

Eqs.~\eqref{eq:ki0_shifted_final} and \eqref{eq:Ti_shifted} can be recast in matrix form
\begin{equation}
\frac{d}{dt} 
\begin{pmatrix} k_{{up}_i} \\ T_i \end{pmatrix}
= \frac{1}{t}
\begin{pmatrix}
a+ b & b \\
b & b
\end{pmatrix}
\begin{pmatrix} k_{{up}_i} \\ T_i \end{pmatrix},
\label{eq:matrix_form}
\end{equation}

with
\begin{equation}
a = \frac{p_1}{2(1+p_2)}, \quad b = \frac{2p_2}{2+5p_2}.
\label{eq:ab_def}
\end{equation}

The exponent characterizing the asymptotic behaviour is given by the largest eigenvalue of the matrix. The characteristic equation reads
\begin{equation}
\begin{aligned}
&\lambda^2 - Tr\lambda + Det = 0\\
&\lambda^2 - (a+2b)\lambda + ab = 0,
\end{aligned}
\label{eq:char_poly}
\end{equation}

which gives
\begin{equation}
\lambda = \frac{a+2b \pm \sqrt{a^2 + 4b^2}}{2}.
\label{eq:eigenvalues}
\end{equation}

The largest eigenvalue is
\begin{equation}
\lambda_{\max} = \frac{1}{2} \left[ \frac{p_1}{2(1+p_2)} + \frac{4p_2}{2+5p_2} + \sqrt{ \frac{p_1^2}{4(1+p_2)^2} + \frac{16p_2^2}{(2+5p_2)^2} } \right],
\label{eq:lambda_max}
\end{equation}

with $p_1 + p_2 = 1$. For $p_2 \ll 1$, we can expand it as
\begin{equation}
\lambda_{\max} = \frac{1}{2} + \frac{p_2^2}{2} + o(p_2^2),
\end{equation}
showing continuity as $p_2 \to 0$ (see Fig. \ref{fig:beta_p1_p2_epsilon}).

\subsubsection{$k_{down} \to \varepsilon$}
The same computation shown in Eq. \eqref{eq:deg0_shifted} can be performed by substituting the down degree with a small quantity that we can then let go to zero, namely $k_{down} \to \varepsilon$, in order to recover the result shown in Sec. \ref{sec:p1-p2 case}. In particular we pose now

\begin{equation}
\frac{d k_{{up}_i}^{(0)}}{dt} = p_1 \frac{k_{{up}_i}^{(0)}}{\sum_j k_{{up}_ji}^{(0)}} + p_2 \sum_{e \ni i} \frac{k_{{up}_e}^{(1)} + \varepsilon}{\sum_e (k_{{up}_e}^{(1)}+\varepsilon)}.
\label{eq:deg0_shifted_epsilon}
\end{equation}

By following the passages carried out in Appendix~ \ref{sec:Lower degree included} we obtain

\begin{equation}
\frac{d}{dt} 
\begin{pmatrix} k_{{up}_i} \\ T_i \end{pmatrix}
= \frac{1}{t}
\begin{pmatrix}
a+ \varepsilon b & 2b \\
\varepsilon b & 2b
\end{pmatrix}
\begin{pmatrix} k_{{up}_i} \\ T_i \end{pmatrix},
\label{eq:matrix_form_epsilon}
\end{equation}
where now
\begin{equation}
a = \frac{p_1}{2(1+p_2)}, \quad b = \frac{p_2}{\varepsilon p_1 +(3+2\varepsilon)p_2}.
\label{eq:ab_def_epsilon}
\end{equation}

As before, we can expand the largest eigenvalues of the characteristic polynomial with $p_1+p_2=1$
\begin{equation}
\lambda = \frac{a+b(2+\varepsilon) \pm \sqrt{(a+b(2+\varepsilon))^2 - 8ab}}{2},
\label{eq:eigenvalues_epsilon}
\end{equation}

getting to
\begin{equation}
\lambda_{\max} = \frac{1}{2} + \frac{p_2^2}{\varepsilon} + o(p_2^2),
\end{equation}
showing continuity in the limit $p_2 \to 0$ for finite values of $\varepsilon$ and a discontinuity if $\varepsilon=0$.

\begin{figure}[H]
    \centering
    \includegraphics[width=0.8\linewidth]{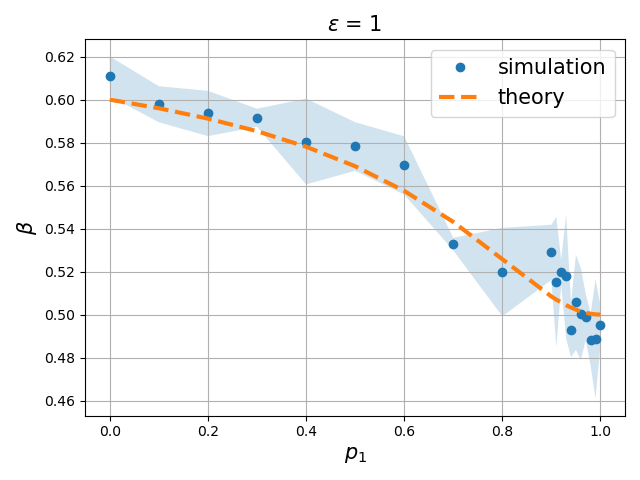}
    \caption{Plot of $\beta_0$ as a function of $p_1$, with $p_1+p_2=1$ and $\varepsilon=1$. The blue dots and the shaded area indicate the simulation results with errors, while the orange dashed line corresponds to the maximum eigenvalue $\lambda$ obtained from Eq.~\eqref{eq:eigenvalues_epsilon}. For numerical convenience, since $\varepsilon>0$, at each time step we allow a newly entering node to form $m=3$ new structures according to the generative rule of the proposed algorithm,}
    \label{fig:beta_p1_p2_epsilon}
\end{figure}

The comparison among the theory and the numerical simulations of the growing parameter $\beta_0$ is reported in Fig. \ref{fig:beta_p1_p2_epsilon}.

\bibliographystyle{unsrt}
\bibliography{references}

\end{document}